\documentclass[a4paper,11pt]{article}
\usepackage{amsmath}
\usepackage{amsfonts}
\usepackage{amssymb}
\usepackage{latexsym}
\usepackage{epsfig}
\usepackage{graphicx}
\usepackage{oldgerm}
\usepackage{theorem}

\def\N{\mathbb{N}}
\def\Z{\mathbb{Z}}

\def\R{\mathbb{R}}

\def\x{{\bf x}}
\def\y{{\bf y}}
\def\z{{\bf z}}
\def\e{{\bf e}}
\def\h{{\bf h}}
\def\v{{\bf v}}
\def\w{{\bf w}}

\def\r{{\bf r}}
\def\n{{\bf n}}
\def\m{{\bf m}}
\def\q{{\bf q}}
\def\veta{\mib {\eta}}
\def\vtheta{\mib {\theta}}
\def\p{{\bf p}}
\def\c{{\bf c}}

\def\1{{\bf 1}}

\def\cS{{\cal S}}
\def\cX{{\cal X}}
\def\cR{{\cal R}}
\def\cV{{\cal V}}
\def\cU{{\cal U}}
\def\cA{{\cal A}}
\def\cT{{\cal T}}
\def\cG{{\cal G}}
\def\cB{{\cal B}}
\def\cF{{\cal F}}
\def\cO{{\cal O}}
\def\cH{{\cal H}}

\def\P{{\mathbb P}}

\def\rC{{\rm C}}

\def\bP{{\bf P}}
\def\bE{{\bf E}}

\def\sa{{\sf a}} 
\def\st{{\sf t}} 
\def\sd{{\sf d}} 

\theorembodyfont{\itshape}

\newtheorem{thm}{Theorem}[section]
\newtheorem{lem}[thm]{Lemma}

\newtheorem{prop}[thm]{Proposition}

\newtheorem{df}[thm]{Definition}

\newcommand{\mib}[1]{\mbox{\boldmath $#1$}}
\newcommand{\SSC}[1]{\section{#1}\setcounter{equation}{0}}
\newcommand{\qed}{\hbox{\rule[-2pt]{3pt}{6pt}}}

\begin{document}

\title{\bf 
Dissipative Abelian Sandpile Models
}
\author{
Makoto Katori
\footnote{
Department of Physics,
Faculty of Science and Engineering,
Chuo University, 
Kasuga, Bunkyo-ku, Tokyo 112-8551, Japan;
e-mail: katori@phys.chuo-u.ac.jp
}}
\date{2 May 2015}
\pagestyle{plain}
\maketitle
\begin{abstract}
We introduce a family of abelian sandpile models with
two parameters $n, m \in \N$
defined on finite lattices on $d$-dimensional torus.
Sites with $2dn+m$ or more grains of sand are unstable and topple,
and in each toppling 
$m$ grains dissipate from the system.
Because of dissipation in bulk, the models are
well-defined on the shift-invariant lattices 
and the infinite-volume limit of systems can be taken.
From the determinantal expressions, 
we obtain the asymptotic forms of the avalanche
propagators and the height-$(0,0)$ correlations of sandpiles
for large distances in the infinite-volume limit 
in any dimensions $d \geq 2$. 
We show that both of them decay exponentially with
the correlation length
$$
\xi(d, a)=(\sqrt{d} \sinh^{-1} \sqrt{a(a+2)} \ )^{-1}, 
$$
if the dissipation rate $a =m/(2dn)$ is positive.
By considering a series of models with increasing $n$, 
we discuss the limit $a \downarrow 0$
and the critical exponent defined by
$\nu_{a}=- \lim_{a \downarrow 0} \log \xi(d, a)/
\log a$ is determined as 
$$
\nu_{a}=1/2
$$ for all $d \geq 2$. 
Comparison with the $q \downarrow 0$ limit
of $q$-state Potts model in external magnetic field 
is discussed.
\end{abstract}


\tableofcontents

\SSC{Introduction \label{sec:introduction}}

Let $d \in \{2,3, \dots\}$ and $L \in \N \equiv \{1,2,3, \dots\}$.
Consider a box in the $d$-dimensional hypercubic 
lattice $B_L=\{-L, -L+1, \dots, L\}^{d} \subset \Z^{d}$, where
$\Z$ denotes the collection of all integers.
We impose {\it periodic boundary conditions}
for all $d$ directions 
and obtain a lattice on a torus (toroidal), 
which is denoted by $\Lambda_L$.
The number of sites in $\Lambda_L$ is given by
$|\Lambda_L| = (2L+1)^d$.
In the present paper
we study a family of Markov processes on $\Lambda_{L}$, 
$h_t =\{h_t(\z)\}_{\z \in \Lambda_L}$, 
with discrete-time $t \in \N_0 \equiv \{0\} \cup \N$.

Assume $n, m \in \N$ and let 
$$
  a=\frac{m}{2dn}
 \qquad \hbox{and} \qquad
 h_{\rm c}=2d (1+a).
$$
Define a real symmetric matrix with size $(2L+1)^{d}$, 
\begin{eqnarray}
\Delta_{L}(\x, \y)=
\left\{
   \begin{array}{rl}
      h_{\rm c}, 
 & \quad \hbox{if} \quad \x=\y, \\
      -1, & \quad \hbox{if} \quad |\x-\y| = 1, \\
      0, & \quad \hbox{otherwise}, \\
   \end{array}\right.
\label{eqn:Delta}
\end{eqnarray}
where $\x =(x_1, \dots, x_d), \y=(y_1, \dots, y_d) \in \Lambda_L$
and $|\x-\y|=\sqrt{\sum_{i=1}^{d}(x_{i}-y_{i})^{2}}$. 
Let $\1(\omega)$ be the indicator function
of an event $\omega$; 
${\bf 1}(\omega)=1$, if $\omega$ occurs
and ${\bf 1}(\omega)=0$, otherwise.
The configuration space is
$$
\cS_L
= \left\{ 0, \frac{1}{n}, \frac{2}{n}, \dots, h_{\rm c}-\frac{1}{n} \right\}^{\Lambda_L}.
$$
Given a configuration $h_t \in \cS_L, t \in \N_0$, 
$h_{t+1} \in \cS_L$ is determined by the following algorithm.

\begin{description}
\item{(i)} \ Choose one site in $\Lambda_{L}$ at random.
Let $\x$ be the chosen site
and define 
$$
\eta^{\x}_{(1)}(\z)=h_{t}(\z)+ \frac{1}{n} \1(\z=\x), \quad \z \in \Lambda_L.
$$
If $\eta^{\x}_{(1)}(\x) < h_{\rm c}$, then
$\eta^{\x}_{(1)} \equiv \{\eta^{\x}_{(1)}(\z)\}_{\z \in \Lambda_L} \in \cS_{L}$.
In this case, we set
$h_{t+1}=\eta^{\x}_{(1)}$.
\item{(ii)} \ 
If $\eta^{\x}_{(1)}(\x) = h_{\rm c}$, then
$\eta^{\x}_{(1)} \notin \cS_{L}$.
In this case, we consider a finite series of configurations 
$\{\eta^{\x}_{(1)}, \cdots, \eta^{\x}_{(\tau)} \}$ with $\exists \tau \in \N$  
recursively as follows.
Assume that $\eta^{\x}_{(\ell)} \notin \cS_{L}$ with $\ell \geq 1$,
then 
$A^{\x}_{(\ell)}(h_{t}) \equiv 
\{ \z \in \Lambda_{L}: \eta^{\x}_{(\ell)}(\z) \geq h_{\rm c} \}
\not= \emptyset$ and define
$$
\eta^{\x}_{(\ell+1)}(\z)=\eta^{\x}_{(\ell)}(\z)
-\sum_{\y: \y \in A^{\x}_{(\ell)}(h_{t})}\Delta_{L}(\y, \z),
\quad \z \in \Lambda_L.
$$
If $\eta^{\x}_{(\ell+1)} \in \cS_{L}$, then $\tau=\ell+1$
and $h_{t+1}=\eta^{\x}_{(\tau)}$.
Remark that $\tau=\tau(\x, h_t)$ and 
$\tau < \infty$ by
$\sum_{\z: \z \in \Lambda_{L}} \Delta_{L}(\y, \z) > 0, \forall \y \in \Lambda_{L}$
as explained below.
\end{description}

\begin{figure}
\hskip 4cm
\includegraphics[width=0.4\linewidth]{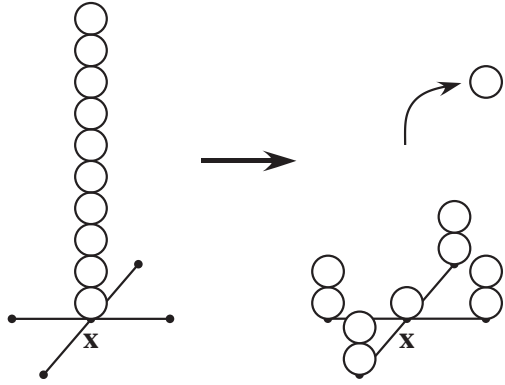}
\caption{A toppling for the DASM with the parameters
$d=2, n=2$ and $m=1$. 
In this case $h_{\rm c}=2dn+m=9$, and thus the site $\x$
with height $h(\x)=10$ is unstable.
In a toppling, $h_{\rm c}=9$ grains of sand drop from the site $\x$, in which
$n=2$ grains land on each nearest-neighbor site,
$m=1$ grain is dissipated from the system, 
while $h(\x)-h_{\rm c}=1$ grain remains on the site $\x$.
}
\label{fig:toppling}
\end{figure}

We think that $1/n$ is a unit of grain of sand and
$h_t(\z) n $ represents the height of sandpile at site $\z$
measured in this unit. 
The step (i) simulates a random deposit of a grain of sand.
In the step (ii), for each $1 \leq \ell \leq \tau$, the sites
$\y \in A^{\x}_{(\ell)}(h_{t})$ are regarded as unstable sites and
the process 
$$
\{\eta^{\x}_{(\ell)}(\z) \}_{\z \in \Lambda_L} \to 
\{\eta^{\x}_{(\ell)}(\z)-\Delta_{L}(\y, \z)\}_{\z \in \Lambda_L},
$$
is called a {\it toppling} of the site $\y$
such that 
$$
\mbox{$\Delta_{L}(\y, \y) n=h_{\rm c} n =2dn+m$ grains of sand
drop from the unstable site $\y$}
$$
and 
$$ \mbox{$|\Delta_{L}(\y, \z)|n = n$ grains of sand 
land on each nearest-neighbor site $\z, |\x-\z|=1$}.
$$
Since there are $2d$ nearest-neighbor sites of each site,
$m$ grains are annihilated in a toppling. (See Fig.\ref{fig:toppling}.)
The total number of grains on $\Lambda_L$ decreases
in each toppling and it guarantees $\tau < \infty$.
The configuration space $\cS_{L}$ is a set of all stable configurations
of sandpiles in which height of sandpile is less than 
the threshold value $h_{\rm c}$ at every site;
$h(\z) < h_{\rm c}, \forall \z \in \Lambda_L$.
From a stable configuration $h_{t}$ to another stable configuration
$h_{t+1}$, $\sum_{\ell=1}^{\tau-1} 
|A^{\x}_{(\ell)}(h_{t})|$ 
topplings occur.
Such a series of toppling is called an {\it avalanche}.
(Note that, if $\tau=1$, toppling does not occur.
Even in such a case, we call the transition
from $h_t$ to $h_{t+1}$ an avalanche, which is
just a random deposit of a grain of sand.) 
Define
\begin{equation}
  T(\x, \y, h)= \sum_{\ell=1}^{\tau(\x, h)-1} 
{\bf 1} (\y \in A^{\x}_{(\ell)}(h)),
\quad \x, \y \in \Lambda_L, \quad h \in \cS_L.
\label{eqn:Txyh}
\end{equation}
This is the number of topplings at site $\y \in \Lambda_L$ in an
avalanche caused by a deposit of a grain of sand
at a site $\x \in \Lambda_L$ in the configuration $h \in \cS_L$.

We have assumed that $n, m \in \N$ in the above definition of processes.
If we set $n=1, m=0$, however, we have $a=0$ and
$\Delta_L|_{a=0}$ gives the `rule matrix' of 
the sandpile model introduced by
Bak, Tang and Wiesenfeld (BTW) \cite{BTW87,BTW88}. 
The BTW model have been studied on
finite lattices with {\it open boundary conditions} 
in order to make $\tau$ be finite.
For example, the BTW model is considered on a box $B_L$.
The boundary of box $B_L$ is given by
$\partial B_L =\{\y=(y_{1}, \cdots, y_{d}) \in B_{L}:
1 \leq \exists i \leq d \ \hbox{s.t.} \
y_{i}=-L \ \hbox{or} \ L \}$.
In the BTW model defined on $B_L$, 
$\sum_{\z: \z \in \Lambda_L}\Delta_{L}|_{a=0}(\y ,\z)=0$
if $\y \in B_{L} \setminus \partial B_{L}$; 
that is, the number of grains of sand is conserved in any
toppling in the bulk of system. 
By imposing the open boundary condition, 
we have 
$\sum_{\z: \z \in \Lambda_L} \Delta_L|_{a=0}(\y, \z) >0$
for $\y \in \partial B_L$ and dissipation of grains of sand
can occur in topplings at the boundary sites.
In the present model, in every toppling at any site $\y \in \Lambda_L$, 
$\sum_{\z: \z \in \Lambda_{L}} \Delta_{L}(\y, \z) n = m$
grains of sand dissipate from the system and hence
$\tau < \infty$ is guaranteed in the shift-invariant system.
The quantity $a$ indicates the rate of dissipation in
a toppling. 

The present process belongs to the
class of {\it abelian sandpile models} (ASM) studied by
Dhar \cite{D90}.  We define the operators
$\{\sa(\x)\}_{\x \in \Lambda_L}$ following Dhar by
$$
h_{t+1}=\sa(\x) h_{t}, \quad \x \in \Lambda_L, 
$$
where $h_{t}, h_{t+1} \in \cS_{L}$ and
the site $\x$ is the chosen site in the first
step (i) of the algorithm at time $t$.
That is, $\sa(\x)$ represents an avalanche caused by a deposit
of a grain of sand at $\x$.
Then the above algorithm guarantees the
{\it abelian property} of avalanches (see Lemma \ref{thm:abelian}
in Section \ref{sec:abelian})
\begin{equation}
  [\sa(\x), \sa(\y)] \equiv \sa(\x) \sa(\y) - \sa(\y) \sa(\x) = 0, \quad
\forall \x, \y \in \Lambda_{L}.
\label{eqn:abelian1}
\end{equation}

We call the present Markov process the $d$-dimensional
{\it dissipative abelian sandpile model} 
(DASM for short).
The two-dimensional case was
studied numerically \cite{GLJ97}
and analytically \cite{VZ98,TK00,MR01}.
In the present paper, we will discuss the models
in general dimensions $d \geq 2$ in finite and infinite lattices.
See also \cite{VD01}.
As shown in \cite{MRS04,SV09,JRS14} 
the DASM is useful to construct the infinite-volume limit of
avalanche models.
Importance of the abelian sandpile models in the extensive study
of {\it self-organized criticality} in the statistical mechanics and
related fields is discussed in \cite{Pru12}.

\SSC{Basic Properties of Dissipative Abelian Sandpile Model \label{sec:basic}}
\subsection{Abelian property \label{sec:abelian}}

First we prove the abelian property of avalanches (\ref{eqn:abelian1}).

\begin{lem}[Dhar \cite{D90}]
\label{thm:abelian}
Assume that the avalanche operators $\{\sa(\x)\}_{\x \in \Lambda_L}$
act on $\cS_L$. Then
$$
[\sa(\x), \sa(\y)] =0, \quad \forall \x, \y \in \Lambda_L.
$$
\end{lem}
\noindent{\it Proof.} 
Let $\cX_{L}=\Z^{\Lambda_{L}}$.
Define three sets of maps 
from $\cX_{L}$ to $\cX_{L}$; 
$\{\tilde{\st}(\x)\}_{\x \in \Lambda_{L}}$,
$\{\st(\x)\}_{\x \in \Lambda_{L} }$ and 
$\{\sd(\x)\}_{\x \in \Lambda_{L}}$
as follows.
For $\x \in \Lambda_{L}$ and $\eta=\{\eta(\x)\}_{\x \in \Lambda_L} \in \cX_{L}$ define
\begin{eqnarray}
  \tilde{\st}(\x) \eta(\z) &=& \eta(\z) - \Delta_{L}(\x, \z), \nonumber\\
  \st(\x) \eta(\z) &=&
\left\{
   \begin{array}{ll}
      \eta(\z)-\Delta_{L}(\x,\z),
 & \quad \hbox{if} \quad \eta(\x) \geq h_{\rm c}, \\
      \eta(\z), & \quad \hbox{otherwise,} \\
   \end{array}\right. \nonumber\\
 \sd(\x) \eta(\z) &=& \eta(\z) +\frac{1}{n} \1(\z=\x), \qquad \z \in \Lambda_L. \nonumber
\end{eqnarray}
By definition of $\tilde{\st}$,
$$
  \tilde{\st}(\y) \tilde{\st}(\x) \eta(\z)=\eta(\z)-\Delta_L(\x,\z)-\Delta_L(\y,\z), 
  \quad \z \in \Lambda_{L}.
$$
Similarly we have
$$
  \tilde{\st}(\x) \tilde{\st}(\y) \eta(\z)=\eta(\z)-\Delta_L(\y,\z)-\Delta_L(\x,\z), 
  \quad \z \in \Lambda_{L}.
$$
Therefore $\tilde{\st}(\y) \tilde{\st}(\x) \eta=\tilde{\st}(\x) \tilde{\st}(\y) \eta, \forall \eta \in \cX_{L}$,
that is
\begin{equation}
  [\tilde{\st}(\x), \tilde{\st}(\y)]=0, \quad
  \forall \x, \y \in \Lambda_{L}.
\label{eqn:tilde_t}
\end{equation}
Assume that $\y \not= \x$. Then
$$
\tilde{\st}(\y) \eta(\x)=\eta(\x)-\Delta(\y,\x)=
  \left\{
   \begin{array}{ll}
      \eta(\x)+1,
 & \quad \hbox{if} \quad |\x-\y|=1, \\
      \eta(\x), & \quad \hbox{if} \quad |\x-\y| >1. \\
   \end{array}\right. 
$$
It implies that
if $\eta(\x) \geq h_{\rm c}$ then 
$\tilde{\st}(\y) \eta(\x) \geq h_{\rm c}, \forall \y \not= \x$, that is, 
any site cannot be stabilized by topplings
which occur at other sites.
Therefore, the definition of $\st(\x)$ and (\ref{eqn:tilde_t})
give
\begin{equation}
 [\st(\x), \st(\y)] = 0, \quad
 \forall \x, \y \in \Lambda_{L}.
\label{eqn:t}
\end{equation}
It is obvious that
\begin{equation}
  [\st(\x), \sd(\y)]=0, \quad
  \forall \x, \y \in \Lambda_{L}.
\label{eqn:d}
\end{equation}
Consider the situation that $h \in \cS_L$ and 
$A^{\x}_{(\ell)}(h) \not= \emptyset$,
$1 \leq \ell \leq \tau$.
By (\ref{eqn:t}), 
$\prod_{\z: \z \in A^{\x}_{(\ell)}(h)} \st(\z)$
is independent of the order of the products
of $\st(\z)$'s. Then we can write
$$
  \sa(\x) h = \left[ \prod_{\ell=1}^{\tau-1}
  \left( \prod_{\z: \z \in A^{\x}_{(\ell)}(h)} 
  \st(\z) \right) \right] \sd(\x) h, \quad \x \in \Lambda_L, \quad h \in \cS_L.
$$
By (\ref{eqn:t}) and (\ref{eqn:d}), the lemma is proved. \qed
\vskip 0.3cm

\subsection{Recurrent configurations \label{sec:recurrent}}

Consider a subset of $\cS_L$ defined by
$$
\cR_{L}=\{h \in \cS_{L}: \ \forall \x \in \Lambda_{L}, 
\exists k(\x) \in \N, \ \mbox{s.t.} \
(\sa(\x))^{k(\x)}h = h \},
$$
which is called the set of {\it recurrent configurations}.

\begin{figure}[htbp]
\hskip 5.5cm
\includegraphics[width=0.3\linewidth]{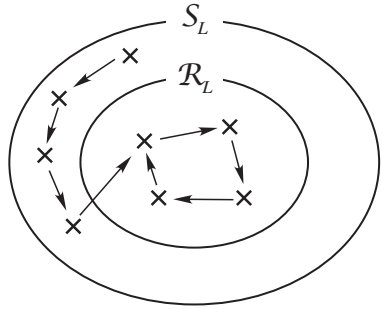}
\caption{The set of recurrent configurations $\cR_L$
is closed under avalanches.
}
\label{fig:recurrent}
\end{figure}

\begin{lem}[Dhar \cite{D90}]
\label{thm:recurrent}
If $h \in \cR_{L}$, then $\sa(\x) h \in \cR_{L}$
for any $\x \in \Lambda_{L}$.
That is, $\cR_{L}$ is closed under avalanches
(see Fig.\ref{fig:recurrent}). 
\end{lem}
\noindent{\it Proof.} 
By definition, if $h \in \cR_{L}$, then for any $\y \in \Lambda_{L}$, 
$\exists k(\y) \in \N$, s.t. $(\sa(\y))^{k(\y)} h=h$.
If we operate $\sa(\x), \x \in \Lambda_L$ on the both sides of this equation,
then we have
$\sa (\x) (\sa(\y))^{k(\y)} h = \sa(\x) h$.
By Lemma \ref{thm:abelian},
LHS$=(\sa(\y))^{k(\y)} \sa(\x) h$. This equality
implies that $\sa(\x) h \in \cR_{L}$.
Since it is valid for
any $\x \in \Lambda_{L}$, the proof is completed. \qed
\vskip 0.3cm

Consider a $(2L+1)^d$-dimensional vector space $\cV_L$, 
in which the orthonormal basis is 
given by $\{\e(\z)\}_{\z \in \Lambda_L}$.
For each configuration $\eta \in \cX_L $, we assign a vector
\begin{equation}
\veta = \sum_{\z: \z \in \Lambda_L}
\eta(\z) \e(\z)
=\sum_{\z: \z \in \Lambda_L} n \eta(\z) \frac{\e(\z)}{n},
\label{eqn:veta1}
\end{equation}
where $1/n$ denotes the unit of grain of sand.
Assume that $h \in \cR_L$; 
for each $\x \in \Lambda_L$, 
there is $k(\x) \in \N$ such that
\begin{equation}
  (\sa(\x))^{k(\x)} h=h.
\label{eqn:recb1}
\end{equation}
Consider the vector corresponding to the 
configuration $(\sd(\x))^{k(\x)} h$, 
\begin{equation}
\veta=\left( h(\x)+\frac{k(\x)}{n} \right) \e(\x)
+ \sum_{\z: \z \not= \x} h(\z) \e(\z) \in \cV_L.
\label{eqn:vetap}
\end{equation}
Then (\ref{eqn:recb1}) claims that there exists a set
$\{r(\z) \in \N: \z \in \Lambda_L\}$ such that
\begin{equation}
  \h=\veta+\sum_{\z: \z \in \Lambda_L}
  \left( \sum_{\y: \y \in \Lambda_L} r(\y) \Delta_L(\y, \z) \right) \e(\z).
  \label{eqn:recb2}
\end{equation}
Note that (\ref{eqn:recb2}) is written as
$$
  \h=\veta+\sum_{\y: \y \in \Lambda_L}
  r(\y) \v(\y)
$$
with
\begin{equation}
  \v(\x)= \sum_{\z: \z \in \Lambda_{L}} \Delta_{L}(\x, \z) \e(\z),
  \quad \x \in \Lambda_L.
\label{eqn:v}
\end{equation}

\begin{figure}[htbp]
\hskip 5.5cm
\includegraphics[width=0.3\linewidth]{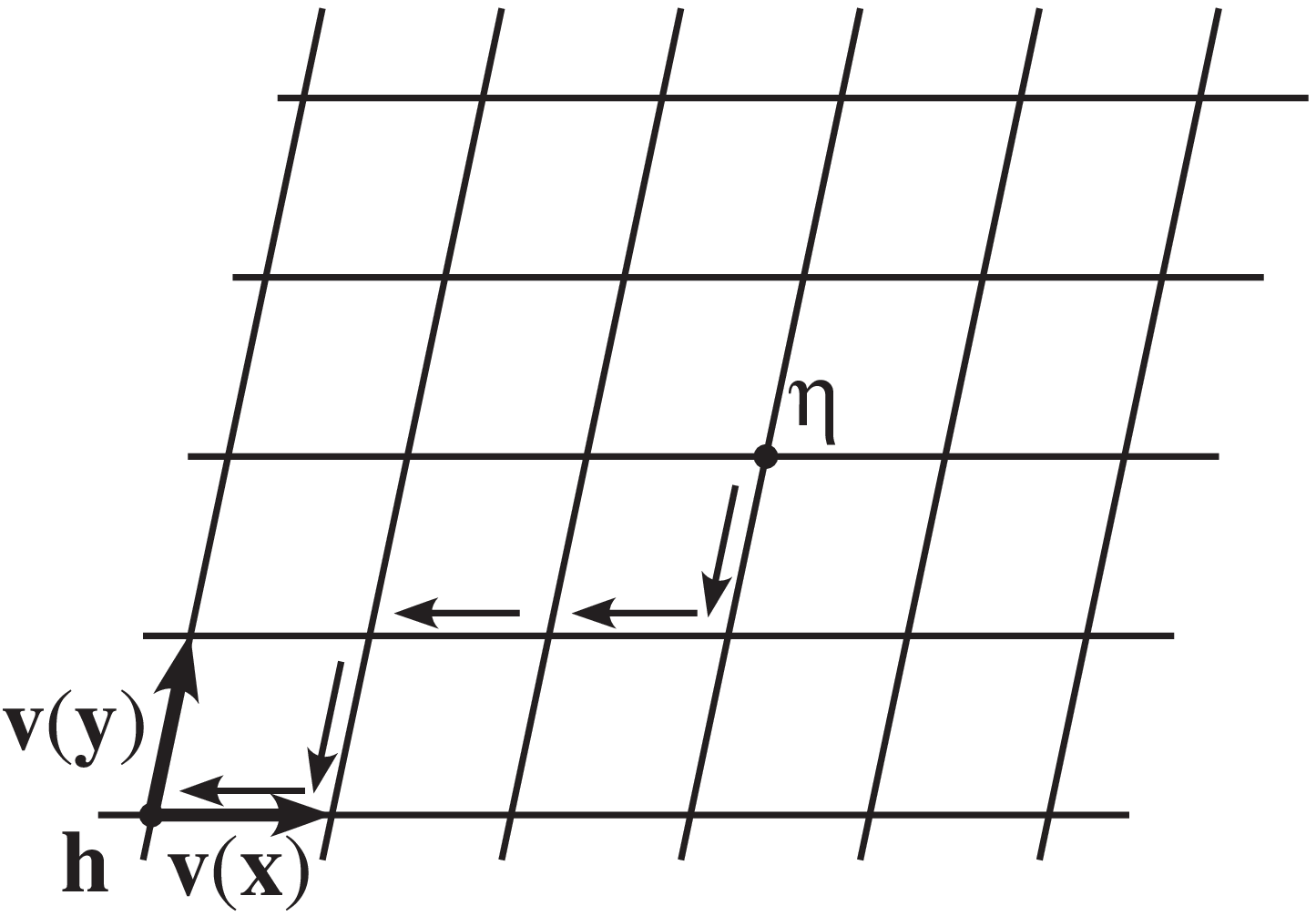}
\caption{Hypercubic lattice $\Omega$
with the basis $\{\v(\x)\}_{\x \in \Lambda_L}$ in $\cV_L$.
Every avalanche from an unstable configuration
$\veta$ given by (\ref{eqn:vetap}) to 
a recurrent configuration $h \in \cR_L$ is represented
by a lattice path $\veta \leadsto \h$ on $\Omega$.
}
\label{fig:lattice1}
\end{figure}

We can say that, given $h \in \cR_L$, all points $\{\veta \}$
given by (\ref{eqn:vetap}) are
identified with sites of a hypercubic lattice $\Omega$
with the basis $\{\v(\x)\}_{\x \in \Lambda_L}$ in $\cV_L$. 
(See Fig.\ref{fig:lattice1}.) 
Consider a primitive cell (fundamental domain) of the lattice defined by
\begin{equation}
\cU_L=\left\{\sum_{\x: \x \in \Lambda_L} \c(\x) \v(\x) :
0 \leq \c(\x) < 1, \x \in \Lambda_L \right\}
\subset \cV_L.
\label{eqn:UL}
\end{equation}
By definition, the intersection of the lattice $\Omega$ and $\cU_L$
is a singleton, say $\p$. We assume that
the origin of this lattice is given by $\p$
and express the lattice by $\Omega^{\p}$.
We consider a collection of all lattices with the same 
basis (\ref{eqn:v}) having distinct origin in $\cU_L$,
$\{\Omega^{\p} : \p \in \cU_L \}$.
Then there establishes a bijection between $\cR_L=\{h\}$
and $\{\Omega^{\p} : \p \in \cU_L \}$.

\begin{lem}[Dhar \cite{D90}]
\label{thm:det_L}
The number of recurrent configuration is given by
$$
|\cR_L| = n^{(2L+1)^d} \det \Delta_L.
$$
\end{lem}
\noindent{\it Proof.} 
The above bijection implies
$|\cR_L|=|\{\Omega^{\p} : \p \in \cU_L \}|$.
Since the unit of grain of sand is $1/n$, 
the origins $\{\p \}$ of lattices $\{\Omega^{\p}\}$ should be 
in $(\Z/n)^{\Lambda_L}$, and hence
$$
|\{\Omega^{\p} : \p \in \cU_L \}|
=\Big| \cU_L \cap (\Z/n)^{\Lambda_L} \Big|
=n^{(2L+1)^d} \times
\mbox{(the volume of $\cU_L$)}.
$$
The volume of $\cU_L$ given by (\ref{eqn:UL}) with (\ref{eqn:v})
is $\det \Delta_L$ and the proof is completed. \qed
\vskip 0.3cm
\begin{figure}[htbp]
\hskip 5.5cm
\includegraphics[width=0.4\linewidth]{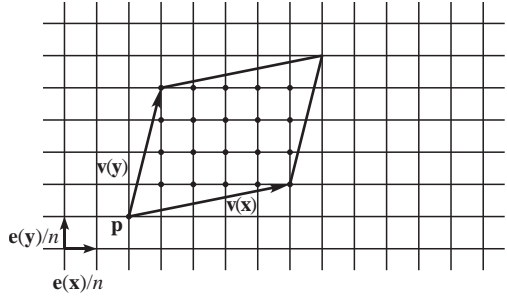}
\caption{A primitive cell of $\Omega$ on the lattice
$(\Z/n)^{\Lambda_L}$. 
Since the unit of grain of sand is $1/n$, 
the origin $\p$ of lattice $\Omega$ should be 
at a site of $(\Z/n)^{\Lambda_L}$.
}
\label{fig:lattice2}
\end{figure}

\subsection{Stationary distribution \label{sec:stationary}}

For $h \in \cR_L$, let $\P_{L}^h$ be the probability law of the DASM
starting from the configuration $h_0=h$.

\begin{df}
\label{thm:a_-1}
If we restrict $\{\sa(\x)\}_{\x \in \Lambda_L}$ to $\cR_L$,
inverse of the avalanche operator can be defined by
$$
 \sa(\x)^{-1} =\sa(\x)^{k(\x)-1}, 
\quad \x \in \Lambda_{L}.
$$
\end{df}

Assume that $h \in \cR_L$ is given.
Define
\begin{eqnarray}
&& \mu_t(X)= \P^{h}(h_{t}=X), \nonumber\\
&& W(X \to Y) = \P^{h}(h_{t+1}=Y | h_{t}=X), 
\quad t \in \N_0, \quad X, Y \in \cR_L.
\nonumber
\end{eqnarray}
Consider the Master equation
$$
\mu_{t+1}(X)=\mu_t(X)-\sum_{Y: Y \in \cR_L} \mu_t(X) W(X \to Y)
+ \sum_{Y: Y \in \cR_L}
\mu_t(Y) W(Y \to X),
$$
where we have used the assumption that 
$h_{0}=h \in \cR_L$ and Lemma \ref{thm:recurrent}.
By definition of the DASM, we can find that, for $X, Y \in \cR_L$, 
\begin{eqnarray}
W(X \to Y) &=& \sum_{\x: \x \in \Lambda_L}
{\rm Prob}(\x \ \hbox{is chosen}) \1(\sa(\x) X=Y) \nonumber\\
&=& \frac{1}{|\Lambda_L|} \sum_{\x: \x \in \Lambda_L}
\1(\sa(\x) X = Y) \nonumber\\
&=& \frac{1}{(2L+1)^d} \sum_{\x: \x \in \Lambda_L}
\1(X = \sa^{-1}(\x) Y). \nonumber
\end{eqnarray}
Then we have
$$
\mu_{t+1}(X)-\mu_t(X)
=\frac{1}{(2L+1)^d} \sum_{\x: \x \in \Lambda_L}
\{ \mu_t(\sa(\x)^{-1} X)-\mu_t(X) \}, 
\quad \forall X \in \cR_L.
$$
It implies that the uniform measure on $\cR_L$,
$$
\mu(X)=\frac{1}{|\cR_L|} \1(X \in \cR_L)
= \frac{1}{n^{(2L+1)^d} \det \Delta_L} \1(X \in \cR_L), \quad X \in \cX_L
$$
is a stationary distribution of the process.

\begin{lem}
\label{thm:irreducible}
The DASM on $\Lambda_L$ is irreducible on $\cR_L$.
\end{lem}
\noindent{\it Proof.} 
Consider the configuration $\overline{h} \in \cS_L$, such that
$\overline{h}(\x)=h_{\rm c}-1/n, \forall \x \in \Lambda_L$.
Now we take two arbitrary configurations $X$ and $Y$ from $\cR_L$.
We have
\begin{equation}
\overline{h}=\prod_{\x: X(\x) < h_{\rm c}-1/n}
(\sa(\x))^{h_{\rm c}-1/n -X(\x)} X
=\prod_{\x: Y(\x) < h_{\rm c}-1/n}
(\sa(\x))^{h_{\rm c}-1/n -Y(\x)} Y.
\label{eqn:reach}
\end{equation}
Since this means that the configuration $\overline{h}$ is
reachable form $X$ and $Y$ by
avalanches, Lemma \ref{thm:recurrent} guarantees 
that $\overline{h} \in \cR_L$.
Since we have assumed that $Y \in \cR_L$,
$(\sa(\x))^{k(\x)} Y=Y$
with some $k(\x) \in \N$ for any $\x \in \Lambda_L$.
Therefore, the second equality of (\ref{eqn:reach})
gives (see Definition \ref{thm:a_-1})
\begin{equation}
Y=\prod_{\x: Y(\x) < h_{\rm c}-1/n}
(\sa(\x))^{k(\x)-(h_{\rm c}-1/n -Y(\x))} \overline{h}.
\label{eqn:reach2}
\end{equation}
Combining (\ref{eqn:reach}) and (\ref{eqn:reach2}) gives
$$
Y=\prod_{\x: Y(\x) < h_{\rm c}-1/n}
(\sa(\x))^{k(\x)-(h_{\rm c}-1/n -Y(\x))}
\prod_{\y: X(\y) < h_{\rm c}-1/n}
(\sa(\y))^{h_{\rm c}-1/n -X(\y)} X.
$$
Let
$\sigma=\sum_{\x: Y(\x) < h_{\rm c}-1/n}
\{k(\x)-(h_{\rm c}-1/n -Y(\x)) \}
+ \sum_{\x: X(\x) < h_{\rm c}-1/n}
\{ h_{\rm c}-1/n -X(\x) \}.
$
Then we see
$$
\P^{h_{0}}(h_{t+s}=Y | h_t=X)
\geq \left(\frac{1}{|\Lambda_{L}|}\right)^{\sigma}
\qquad \hbox{for} \ s \geq \sigma.
$$
Since RHS is strictly positive for finite $L$, this completes
the proof. \qed
\vskip 0.3cm
Then the following is concluded by the general theory of
Markov chains (see, for example, Chapter 6.4 of \cite{GS92}).
\begin{prop}
\label{thm:stationary}
The stationary distribution of the DASM is uniquely given by 
the uniform measure on $\cR_L$.
\end{prop}
We write the probability law of the DASM on $\Lambda_L$ 
in the stationary distribution
as $\bP_L$ and its expectation as $\bE_L$.

\subsection{Allowed configurations and spanning trees \label{sec:allowed}}

Dhar also introduced a subset of $\cS_L$ called
a collection of {\it allowed configurations} $\cA_L$ \cite{D90}.
He defined that for $h \in \cS_L$, 
if there is a subset $F \subset \Lambda_L$ such that
$F \not= \emptyset$ and
\begin{equation}
 h(\y) < \sum_{\x: \x \in F, \x \not= y}
  (-\Delta_L(\x, \y)), 
  \quad \forall \y \in F,
\label{eqn:FSC}
\end{equation}
then $h \in \cS_L$ has a 
{\it forbidden subconfiguration} (FSC) on $F$.
Then define 
$$
   \cA_L=\{h \in \cS_L:
  \mbox{$h$ has no FSC} \}.
$$

\begin{lem} 
\label{thm:RandA}
For the DASM on $\Lambda_L$, 
$$
\cR_L \subset \cA_L.
$$
\end{lem}
\noindent{\it Proof.} 
In the proof of Lemma \ref{thm:irreducible}
we have shown that $\overline{h} \in \cR_L$
and all recurrent stares are reachable from
this configuration $\overline{h}$.
We can prove that $\overline{h} \in \cA_L$ as follows.
We assume that the contrary; 
there exists a finite nonempty set $F \subset \Lambda_L$ 
satisfying (\ref{eqn:FSC}).
In the DASM, however, for any $\y \in F$, 
$\overline{h}(\y)=
h_{\rm c}-1/n=2d+(m-1)/n \geq 2d 
\geq \sum_{\x: \x \in F: \x \not= \y }
(-\Delta_L(\x,\y))$, which contradicts our assumption.
Since both $\cR_L$ and $\cA_L$ include
$\overline{h}$, it is enough to show that $\cA_L$
is closed under the process of avalanche to prove the
lemma, since we have already proved that
$\cR_L$ is so in Lemma \ref{thm:recurrent}.
Remark that addition of particles only increases 
$h$ and such procedure on an allowed configurations cannot create
any FSC. Here we assume that there exists an allowed
configuration $h$ such that by a single toppling
at the site $\x$ it becomes to contain a FSC.
Write 
$
  h^{\prime}=\st(\x) \sd(\x) h$, that is,
\begin{equation}
  h^{\prime}(\y)=h(\y)+\frac{1}{n} \1(\y=\x)-\Delta_L(\x, \y) , 
  \quad \forall \y \in \Lambda_L.
\label{eqn:eq1}
\end{equation}
By assumption, there exists $F \not= \emptyset$ such that
\begin{equation}
  h^{\prime}(\y) < \sum_{\z: \z \in F: \z \not= \y}
  (-\Delta_L(\z, \y)), 
  \quad \forall \y \in F.
\label{eqn:eq2}
\end{equation}
Combining (\ref{eqn:eq1}) and (\ref{eqn:eq2}) gives
$$
  h(\y) < \sum_{\z: \z \in F, \z \not= \y}
  (-\Delta_L(\z, \y)) + \Delta_L(\x, \y), 
  \quad \forall \y \in F \setminus \{\x\}.
$$
Since $\Delta_L(\x, \y) \leq 0$ for $\x \not= \y$,
this inequality means that $h$ has a FSC
on $F \setminus \{\x\}$ and this contradicts our
assumption that $h$ is allowed.
Since any avalanche consists of addition of a particle
and a series of topplings, the proof is completed. \qed
\vskip 0.3cm

\begin{figure}[htbp]
\hskip 5.5cm
\includegraphics[width=0.3\linewidth]{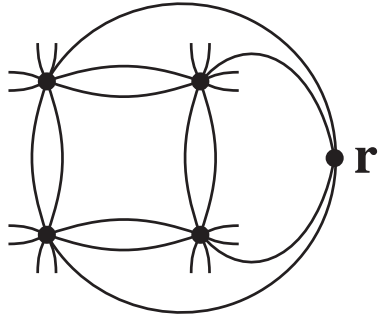}
\caption{
A part of the graph $G_L=(G_L^{(v)}, G_L^{(e)})$ 
associated to the DASM $(\Lambda_L, \Delta_L)$ is
illustrated for the case that
$d=2, n=2$ and $m=1$.
In this case, each pair of
the nearest-neighbor vertices are connected by $n=2$ edges 
and each vertex is connected to the `root' $\r$
by $m=1$ edge.
}
\label{fig:graph_G}
\end{figure}

\begin{df}
\label{thm:graph}
Given a pair $(\Lambda_L, \Delta_L)$, let 
$G_L^{(v)}=\Lambda_L \cup \{\r\}$ with 
an additional vertex $\r$ (the `root'), and
$G_L^{(e)}$ be the collection of 
$|\Delta_L(\x,\y)| n=n$ edges between 
$\x, \y \in \Lambda_L, \x \not=\y$,
and $\sum_{\y: \y \in \Lambda_L} \Delta(\x,\y) n=m$ 
edges between $\x \in \Lambda_L$ and $\r$.
(See Fig.\ref{fig:graph_G}.) 
Graph $G_L$ associated to $(\Lambda_L, \Delta_L)$ is defined as
$$
G_L=(G_L^{(v)}, G_L^{(e)}).
$$
\end{df}

\begin{df}
\label{thm:spanning_tree}
We say a graph $T$ on $G_L$ is a {\it spanning tree},
if the number of vertices of $T$ is $|G_L^{(v)}|=
|\Lambda_L|+1$,
the number of connected components is one, and
the number of loops is zero.
\end{df}
\begin{lem}
\label{thm:spanning_tree2}
Let $
\cT_L=\{ \hbox{\rm spanning tree on $G_L$
associated to $(\Lambda_L, \Delta_L)$} \}$.
Then
$$
  |\cT_L|=n^{(2L+1)^d} \det \Delta_L.
$$
\end{lem}
\noindent{\it Proof.} 
See p.133 of \cite{MD92} and
Theorem 6.3 in \cite{B93}.
\vskip 0.3cm
\begin{lem}[Majumdar and Dhar \cite{MD92}]
\label{thm:AandT}
There establishes a bijection between $\cA_L$
and $\cT_L$.
\end{lem}
\noindent{\it Proof.} 
First we order all edges incident on each
site $\x \in G_L^{(v)}$ in some order of preference.
For each configuration $h \in \cA_L$, 
we consider a following discrete-time growth process of graph
on $G_L$,
which is called a {\it burning process} on $(G_L, h)$.
Let $\tilde{V}_{0}=V_{0}=\{ \r \}$,
$E_{0}=\emptyset$ and $T_{0}=(V_{0}, E_{0})$.
Assume that we have nonempty sets
$T_{t}=(V_{t}, E_{t})$ and $\tilde{V}_{t}$ with 
$t \in \N_0$.
Let
$$
\tilde{V}_{t+1}=\left\{\y \in G_L^{(v)} \setminus V_{t}:
h(\y) \geq \sum_{\x: \x \in  G_L^{(v)} \setminus V_{t}}
(-\Delta_L(\x, \y)) \right\}.
$$
For each $\y \in \tilde{V}_{t+1}$, consider
$$
\tilde{E}_{t+1}(\y)=\Big\{ e \in G_L^{(e)}: e \
\hbox{connects $\y$ and a site in
$\tilde{V}_{t}$}  \Big\}.
$$
We must have
$$
  h(\y) \leq \sum_{\x: \x \in G_L^{(v)} 
  \setminus V_{t}} (- \Delta_L(\x, \y))
  +|\tilde{E}_{t+1}(\y)|,
$$
since $h \in \cS_L$.
If $|\tilde{E}_{t+1}(\y)|=1$, then
name that edge as $e(\y)$.
If $|\tilde{E}_{t+1}(\y)| \geq 2$, then
write 
$$
h(\y) = \sum_{\x: \x \in G_L^{(v)}  \setminus V_{t}} 
(- \Delta_L(\x, \y)) + \frac{s}{n},
$$
and choose the $(s+1)$-th edge in
$\tilde{E}_{t+1}(\y)$ as $e(\y)$.
We define
$$ V_{t+1}=V_{t} \cup \tilde{V}_{t+1}, \quad
E_{t+1}=E_{t} \cup \{e(\y): \y \in \tilde{V}_{t+1} \},
\quad 
\mbox{and} \quad T_{t+1}=(V_{t+1}, E_{t+1}).
$$
By the assumption $h \in \cA_L$, there is a finite
time $\sigma < \infty$ such that
$V_{\sigma}=G_{L}^{(v)}$ and
$E_{\sigma}=G_{L}^{(s)}$.
By the construction,
$T_{\sigma}=(V_{\sigma}, E_{\sigma})$ is a spanning tree on
$G_L$.
Since this growth process of $T_{t},  t \in \{0,1,\cdots, \sigma\}$ is deterministic
for a given configuration $h \in \cA_L$,
it gives an injection from $\cA_L$ to $\cT_L$. 
This fact and Lemma \ref{thm:spanning_tree2} give 
$|\cA_L| \leq |\cT_L|=n^{(2L+1)^d} \det \Delta_L$.
On the other hand, Lemmas \ref{thm:det_L} and \ref{thm:RandA} give
$n^{(2L+1)^d} \det \Delta_L \leq |\cA_L|$.
Then we can conclude $|\cA_L|=n^{(2L+1)^d} \det \Delta_L$
and the burning process gives 
a bijection between $\cA_L$ and $\cT_L$. \qed
\vskip 0.3cm

Combining Lemmas \ref{thm:det_L}, 
\ref{thm:RandA}, \ref{thm:spanning_tree2}, and 
\ref{thm:AandT}, we have the following proposition.
\begin{prop}
\label{thm:RandA_2}
For the DASM on $\Lambda_L$, 
$\cR_L=\cA_L$.
\end{prop}
\SSC{Avalanche Propagators \label{sec:propagator}}
\subsection{Integral expressions for propagators 
\label{sec:integral}}
Define
$$
G_L(\x,\y)=\bE_L[ T(\x, \y, h) ],
\quad \x, \y \in \Lambda,
$$
where $T(\x, \y, h)$ is given by (\ref{eqn:Txyh})
and the expectation is taken over configurations $\{h\}$
in the stationary distribution $\bP_L$. 
$G_L(\x,\y)$ is regarded as the {\it avalanche propagator} from
$\x$ to $\y$ \cite{D90}.
Sometime in an avalanche caused by a deposit of a grain of sand at $\x$,
this site $\x$ topples many times. The set of topplings
between the first and the second toppling at $\x$
is called the first {\it wave} of toppling.
There can occur many waves in one avalanche and
$G_L(\x, \x)$ gives the average number of waves
of topplings in an avalanche \cite{IKP94}.

Consider  the stationary distribution $\bP_L$ of the DASM.
For addition of a particle at any site $\x \in \Lambda_L$,
the averaged influx of grains of sand into a site
$\z \in \Lambda_L$ is given by
$\1(\z=\x)+\sum_{\y: \y \not=\z} G_L(\x, \y)|\Delta_L(\y,\z)|n$,
and  the averaged outflux of them out of $\z$ by
$G_L(\x, \z) \Delta_L(\z,\z)n$ using the avalanche propagators.
In $\bP_L$, equivalence between influx and outflux must hold 
at any site $\z \in \Lambda_L$. This balance equation is written as
$$
  \sum_{\y: \y \in \Lambda_L} G_L(\x, \y) \Delta_L(\y, \z)=
\frac{1}{n} \1(\z=\x) 
\quad \forall \x, \z \in \Lambda_{L}
$$
and thus the propagator is given
using the inverse matrix of $\Delta_L$. 
\begin{lem}[Dhar \cite{D90}]
\label{thm:GL}
\begin{equation}
  G_{L}(\x, \y)= \frac{1}{n} [\Delta_L^{-1}](\x, \y),
\quad \x, \y \in \Lambda_L.
\label{eqn:Green2}
\end{equation}
\end{lem}

The matrix $\Delta_{L}$ can be
diagonalized by the Fourier transformation
from $\x=(x_{1}, \cdots, x_{d})$ to
$\n=(n_{1}, \cdots, n_{d})$,
$$
  U_L(\n,\x)=U_L^{-1}(\x,\n)
  = \frac{1}{(2L+1)^{d/2}}
\exp \left( \frac{2 \pi }{2L+1} \x \cdot \n \right),
$$
where $\x \cdot \n=\sum_{i=1}^d x_i n_i$, 
as
\begin{eqnarray}
&& \sum_{\x: \x \in \Lambda_L} \sum_{\y: \y \in \Lambda_L}
U_L(\n, \x) \Delta_L(\x, \y) U_L^{-1}(\y, \m)
\nonumber\\
&& \qquad = 2d \left\{ (1+a)-\frac{1}{d} \sum_{i=1}^{d} 
\cos \left(\frac{2 \pi}{2L+1} n_{i} \right) \right\} 
\1(\n=\m) \nonumber\\
&& \qquad \equiv \Lambda_{L}(\n,\m),
\quad \n, \m \in \Lambda_L.
\nonumber
\end{eqnarray}
Then, (\ref{eqn:Green2}) is obtained as
\begin{eqnarray}
G_L(\x, \y) &=& \frac{1}{n}
\sum_{\n: \n \in \Lambda_L} \sum_{\m: \m \in \Lambda_L}
U_L^{-1}(\x, \n) [\Delta_L^{-1}](\n,\m)
U_L(\m,\y) \nonumber\\
&=& \frac{1}{2d n} \frac{1}{(2L+1)^d}
\sum_{\n: \n \in \Lambda_L}
\frac{ {\rm e}^{-2 \pi \sqrt{-1}(\x-\y) \cdot \n/(2L+1)}}
{(1+a)-(1/d) \sum_{i=1}^{d} \cos(\frac{2 \pi}{2L+1} n_{i})}.
nonumber
\end{eqnarray}
\begin{lem}
\label{thm:G_infinite}
There exists a limit
$G(\x-\y)=\lim_{L \uparrow \infty} G_L(\x, \y), \x, \y \in \Z^d$ and
\begin{equation}
  G(\x) = \frac{1}{2d n} \prod_{i=1}^d \int_{-\pi}^{\pi} \frac{d \theta_{i}}{2 \pi}
\frac{{\rm e}^{- \sqrt{-1} \x \cdot \vtheta} }
{(1+a) - (1/d) \sum_{i=1}^{d} \cos \theta_{i}}, \quad \x \in \Z^d.
\label{eqn:Green4}
\end{equation}
\end{lem}
\noindent{\it Proof.} 
Consider the Euler-Maclaurin formula
for $f \in \rC^2(\R)$, 
\begin{equation}
  \sum_{n=0}^{M} f(b+n c)=
  \frac{1}{c} \int_{b}^{b+M c} f(\theta) d\theta
  +\frac{1}{2} [f(b)+f(b+M c)]
  +\frac{1}{12} c^2 \sum_{n=0}^{M-1}
  f^{(2)}(b+c(n+\phi)),
\label{eqn:EM}
\end{equation}
where $M \in \N$, $b, c \in \R$, 
$f^{(2)}(\theta)$ is the second derivative of $f(\theta)$, 
and $0 < \phi <1$ (see, for instance, Appendix D in \cite{AAR99}).
Assume that
$$
  f(\theta)=\frac{{\rm e}^{-\sqrt{-1} \alpha_{1} \theta}}
  {(1+a)-(1/d) ( \cos \theta +\alpha_{2} ) },
$$
where $a, \alpha_{1}, \alpha_{2}$ are constants.
Applying the Euler-Maclaurin formula (\ref{eqn:EM})
with $b=-2 \pi L/(2L+1)$, $M=2L$ and
$c=2\pi/(2L+1)$,
we have
\begin{eqnarray}
&& \sum_{n=0}^{2L}
  \frac{{\rm e}^{-2 \pi  \sqrt{-1} \alpha _{1} (n-L)/(2L_{1}+1)}}
  {(1+a)-(1/d)\{ \cos(\frac{2\pi}{2L+1} (n-L) )+\alpha_{2} \} }
\nonumber\\
&& \qquad = (2L+1)
\int_{-2\pi L/(2 L+1)}^{2\pi L/(2L+1)} \frac{d \theta}{2 \pi}
\frac{{\rm e}^{-\sqrt{-1} \alpha_{1} \theta}}
  {(1+a)-(1/d) ( \cos\theta+\alpha_{2} ) } \nonumber\\
&& \qquad \quad + \frac{1}{2} 
  \left[ f \left(-\frac{2\pi L}{2L+1} \right)
+f \left(\frac{2 \pi L}{2L+1} \right) \right]
\nonumber\\
&& \qquad \quad + \frac{1}{12} \left( \frac{2 \pi}{2L+1} \right)^{2}
\sum_{n=0}^{2L-1} 
f^{(2)}\left(\frac{2\pi}{2L+1} (n+\phi-L) 
\right).
\nonumber
\end{eqnarray}
By dividing the both sides of the equality by $2L+1$
and take the limit $L \uparrow \infty$, we obtain
\begin{eqnarray}
&& \lim_{L \uparrow \infty} \frac{1}{2L+1}
\sum_{n=-L}^{L}
  \frac{{\rm e}^{- 2 \pi \sqrt{-1} \alpha_{1} n/(2L_{1}+1)}}
  {(1+a)-(1/d)\{ \cos(\frac{2\pi}{2L+1} n)+\alpha_{2} \} }
\nonumber\\
&& \qquad = 
\int_{-\pi}^{\pi} \frac{d \theta}{2 \pi}
\frac{{\rm e}^{-\sqrt{-1} \alpha_{1} \theta}}
  {(1+a)-(1/d) (\cos\theta+\alpha_{2} ) }.
\nonumber
\end{eqnarray}
Repeating this procedure $d$ times, we can prove
Lemma \ref{thm:G_infinite}.
\qed
\vskip 0.3cm

\subsection{Long-distance asymptotics
\label{sec:long}}
Now we consider the asymptotic form in $|\x| \uparrow \infty$
of $G(\x)$.
Here we follow the calculation found in Section XII.4 of \cite{MW73}
for the asymptotic expansion of two-point spin correlation function
of the two-dimensional Ising model.
By using the identity
$$
  \int_{0}^{\infty} ds {\rm e}^{-\alpha s} 
= \frac{1}{\alpha}
$$
and the definition of the modified Bessel function
of the first kind
$$
  I_{n}(z)= \int_{-\pi}^{\pi} \frac{d \phi}{2 \pi}
{\rm e}^{- \sqrt{-1} n \phi + z \cos \phi},
$$
we have
$$
  G(\x) = \frac{1}{2dn } \int_{0}^{\infty} ds
 {\rm e}^{-(1+a)s} 
 \prod_{i=1}^{d} I_{x_{i}}(s/d).
$$
The asymptotic expansion of $I_{n}(z)$ for large $n$
is found on p.86 in \cite{E53},
$$
I_{n}(z)= \frac{1}{\sqrt{2 \pi}} 
\frac{\exp \left[ (n^2+z^2)^{1/2} 
- n \sinh^{-1}(n/z) \right]}
{(n^2+z^2)^{1/4}} \times \left(1+ \cO(1/n) \right),
$$
and we obtain
\begin{eqnarray}
&& G(\x) = \frac{1}{2d n} 
\left(\frac{1}{2 \pi}\right)^{d/2}
\int_{0}^{\infty} ds 
\prod_{i=1}^{d} \frac{1}{[x_{i}^{2}+(s/d)^2]^{1/4}}
\exp[-g(\x,s)]  \nonumber\\
&& \qquad \qquad \qquad 
\times \left(1+\cO(\max_{i}\{ 1/x_{i}\} )  \right),
\label{eqn:Gxc1}
\end{eqnarray}
where
$$
  g(\x, s)= (1+a)s 
  -\sum_{i=1}^{d} \left[ x_{i}^{2} +\left(\frac{s}{d}\right)^2 \right]^{1/2}
+\sum_{i=1}^{d} x_{i} \sinh^{-1} \left( \frac{d}{s} x_{i} \right).
$$
We can evaluate (\ref{eqn:Gxc1})
by the saddle-point method
and obtain the following result.

\begin{thm}
\label{thm:main1}
Let
\begin{equation}
  c_{1}(d,a) = \frac{1}{4 \pi (a+1)}
  \left[ \frac{\sqrt{a(a+2)d}}{2\pi(a+1)} \right]^{(d-3)/2}
\label{eqn:c1}
\end{equation}
and
\begin{equation}
\xi(d, a)= \frac{1}{\sqrt{d} \sinh^{-1} \sqrt{a(a+2)} }.
\label{eqn:xi}
\end{equation}
Then, for the DASM with $d \geq 2, m, n \in \N, a=m/(2d n)$, 
\begin{equation}
\lim_{r \uparrow \infty} - \frac{1}{r}
\log \left[ \frac{n r^{(d-1)/2} }{c_{1}(d, a)}
 G(\x(r)) \right]
= \frac{1}{\xi(d, a)},
\label{eqn:thm11}
\end{equation}
where 
\begin{equation}
\x(r)=\left ( \frac{r}{\sqrt{d}}, \cdots, \frac{r}{\sqrt{d}} \right) \in \Z^d, \quad r >0.
\label{eqn:xr}
\end{equation}
\end{thm}
\noindent{\it Proof.} 
Let $g^{(1)}(\x, s)$ and $g^{(2)}(\x,s)$ be the first and
second derivatives of $g(\x,s)$ with respect to $s$,
\begin{eqnarray}
  g^{(1)}(\x, s) &=& (1+a)-\frac{1}{d} \sum_{i=1}^{d}
  \left[ 1+ \left(\frac{d}{s} x_{i} \right)^2 \right]^{1/2},
\nonumber\\
  g^{(2)}(\x, s) &=& \frac{d}{s^{3}} \sum_{i=1}^{d}
  x_{i}^{2} \left[ 1+ \left(\frac{d}{s} x_{i} \right)^2
  \right]^{-1/2}.
\nonumber
\end{eqnarray}
For each $\x$, let $s_{0}(\x)$ be the saddle point at which
$g^{(1)}(\x, s)$ vanishes,
\begin{equation}
  g^{(1)}(\x, s_{0}(\x))=0.
  \label{eqn:saddle}
\end{equation}
Then
\begin{eqnarray}
  G(\x) &=& \frac{1}{2dn} \left(\frac{1}{2\pi}\right)^{d/2} 
  \prod_{i=1}^{d} \frac{1}{(x_{i}^2+s_{0}(\x)^2/d^2)^{1/4}}
  \exp[-g(x, s_{0}(\x))] \nonumber\\
&& \qquad \times \int_{-\infty}^{\infty} d u
\exp \left[ -\frac{1}{2} g^{(2)}(\x, s_{0}(\x)) u^2
\right] 
\times \left(1+\cO(\max_{i}\{1/x_{i}\}) \right) \nonumber\\
&=&  \frac{1}{2dn} \left(\frac{1}{2\pi}\right)^{d/2} 
  \prod_{i=1}^{d} \frac{1}{(x_{i}^2+s_{0}(x)^2/d^2)^{1/4}}
  \exp[-g(x, s_{0}(x))] \nonumber\\
&& \qquad \times 
\left(\frac{2 \pi}{g^{(2)}(\x, s_{0}(\x))} \right)^{1/2}
\times \left(1+\cO(\max_{i}\{1/x_{i}\} ) \right).
\nonumber
\end{eqnarray}
Here we can prove that the higher derivatives
of $g(\x,s)$ only give the contributions
of order $\cO(\max_{i} \{1/x_{i}\})$.
See p.304 in \cite{MW73}.
Now we consider the case
$$
   x_{i}=\frac{r}{\sqrt{d}} + \varepsilon_{i},
$$
in which $\varepsilon_i$'s are
finite and fixed and $r \gg 1$.
The equation (\ref{eqn:saddle}) for the saddle point
is now
$$
  \sum_{i=1}^{d} \left( 1+ 
  \frac{d^2}{s_{0}(\x)^{2}}
  \left(\frac{r}{\sqrt{d}}+\varepsilon_{i} \right)^{2}
  \right)^{1/2}
  = (1+a)d,
$$
and it is solved as
$$
  s_{0}(\x)=\sqrt{\frac{d}{a(a+2)}}
  \left( r + \frac{1}{\sqrt{d}} \sum_{i=1}^{d} 
  \varepsilon_{i} + \cO(1/r) \right).
$$
This gives
\begin{eqnarray}
  g(\x, s_{0}(\x)) &=& \sum_{i=1}^{d} \left(
  \frac{r}{\sqrt{d}}+\varepsilon_{i} \right)
  \sinh^{-1} \left[ \frac{d}{s_{0}(x)}
  \left(\frac{r}{\sqrt{d}}+\varepsilon_{i} \right) \right]
  \nonumber\\
  &=& \sqrt{d} r \sinh^{-1} \sqrt{a(a+2)}
  +\sinh^{-1} \sqrt{a(a+2)} \times \sum_{i=1}^{d} \varepsilon_{i}
  +\cO(1/r)
\nonumber
\end{eqnarray}
and
$$
  g^{(2)}(\x, s_{0}(x))
  =\frac{1}{\sqrt{d}} \frac{(a(a+2))^{3/2}}{a+1}
  \frac{1}{r} + \cO(1/r^2).
$$
Then we have the estimation
$$
G(\x) =
\frac{c_1(d,a)}{n}
\frac{1}{r^{(d-1)/2}}
\exp\left[-\frac{r}{\xi(d,a)}
-\lambda(a) \sum_{i=1}^{d} \varepsilon_{i} \right]
\times \left(1+\cO(1/r) \right),
\quad \mbox{as $r \uparrow \infty$}
$$
for $\x=(r/\sqrt{d}+\varepsilon_{1}, \cdots,
r/\sqrt{d}+\varepsilon_{d})$,
where $c_1(d,a)$ and $\xi(d,a)$ are given by 
(\ref{eqn:c1}) and (\ref{eqn:xi}), respectively, and
\begin{eqnarray}
\lambda(a) &\equiv& \frac{\sqrt{d}}{\xi(d,a)} \nonumber\\
&=& \sinh^{-1} \sqrt{a(a+2)} \nonumber\\
&=& \log(1+a+\sqrt{a(a+2)}).
\label{eqn:lambda}
\end{eqnarray}
If we put $\varepsilon_{i}=0, 1 \leq i \leq d$, 
then $G(\x)$ is reduced to be
$$
G(\x(r)) = \bar{G}(r) \times
\left(1+\cO(1/r) \right), \quad
\mbox{as $r \uparrow \infty$}
$$
with
\begin{equation}
  \bar{G}(r) = \frac{c_{1}(d,a)}{n} 
\frac{{\rm e}^{-r/\xi(d,a)}}{r^{(d-1)/2}}. 
\label{eqn:Gbar}
\end{equation}
It proves the theorem. 
\qed

\SSC{Height-$0$ Density and Height-$(0,0)$ Correlations \label{sec:correlation}}

For 
$$
\alpha, \beta \in \left\{0, \frac{1}{n}, \frac{2}{n}, \dots, h_{\rm c}-\frac{1}{n} \right\},
$$
define 
\begin{eqnarray}
P_{\alpha, L}(\x)  &=& \bE_L[ \1(h(\x)=\alpha) ],
\nonumber\\
P_{\alpha \beta, L}(\x, \y) &=& \bE_L [
\1(h(\x)=\alpha) \1(h(\y)=\beta) ],
\quad \x, \y \in \Lambda_L.
\label{eqn:correlation1}
\end{eqnarray}
$P_{\alpha, L}(\x)$ is the probability that the site $\x$ has
the height $\alpha n$ measured in the unit of grain of sand, $1/n$, 
and $P_{\alpha \beta, L}(\x,\y)$ is the 
{\it $(\alpha, \beta)$-height correlation function} 
\cite{MD91,BIP93,P94}.

For the two-dimensional BTW model
on $B_{L}$ with open boundary condition, 
Majumdar and Dhar \cite{MD91} proved the
existence of the infinite-volume limits
\begin{eqnarray}
   P_{0} &=& \lim_{L \uparrow \infty} P_{0, L}(\x), \nonumber\\
   P_{00}(\x(r)) &=& \lim_{L \uparrow \infty} P_{00, L}(0, \x(r)),
\nonumber
\end{eqnarray}
where $\x(r)=(r/\sqrt{2}, r/\sqrt{2})$.
They gave an $8 \times 8$ matrix 
$M_L(r)$, whose elements depend
on $L$ and $r$, such that
$$
  P_{00,L}(0, \x(r))=\det M_L(r), \quad \forall L > \frac{r}{\sqrt{2}},
$$
and showed that every elements
converge in the infinite-volume limit $L \uparrow \infty$
with a finite $r$.
Then the matrix $M(r)=\lim_{L \uparrow \infty} M_L(r)$ is well-defined
and we have the determinantal expression
$$
  P_{00}(\x(r))= \det M(r).
$$
Moreover, they showed that 
$$
  \lim_{r \uparrow \infty} P_{00}(\x(r))=P_{0}^{2},
$$
and
\begin{equation}
C_{00}(\x(r)) \equiv \frac{P_{00}(\x(r))-P_0^2}{P_0^2}
\simeq - \frac{1}{2} r^{-4},
\quad \mbox{as $r \uparrow \infty$}.
\label{eqn:MD2}
\end{equation}
Majumdar and Dhar claimed \cite{MD91} that
the result (\ref{eqn:MD2}) is generalized for the 
$d$-dimensional BTW model with $d \geq 2$ as
\begin{equation}
C_{00}(\x(r)) \sim r^{-2d}, 
\quad \mbox{as $r \uparrow \infty$}.
\label{eqn:MD3}
\end{equation}
In an earlier paper \cite{TK00}, all these facts also
hold for the two-dimensional DASM, if we prepare
$10 \times 10$ matrix $M_L(r)$. (See also \cite{BIP93}
and \cite{P94} for other generalizations of \cite{MD91}.)
Here we show the result for 
the height-0 density and the height-$(0,0)$ correlations
of the DASM with general $d \geq 2$. 

\subsection{Nearest-neighbor correlations \label{sec:nearest}}

First we prove the following Lemma.
\begin{lem}
\label{thm:adjacent}
Any configuration $h \in \cS_L$, in which there are
two adjacent sites $\z_{1}, \z_{2} \in \Lambda_L$,
$|\z_{1}-\z_{2}|=1$, such that
$h(\z_{1}) < 1$ and $h(\z_{2}) < 1$,
is not allowed.
\end{lem}
\noindent{\it Proof.} 
Let $F=\{\z_{1}, \z_{2} \} \subset \Lambda_L$.
Then
$$ 
\sum_{\x: \x \in F, \x \not=\z_{1}} (-\Delta_L(\x, \z_{1}))
=-\Delta_{L}(\z_{2}, \z_{1})=1,
$$
and
$$
\sum_{\x: \x \in F, \x \not=\z_{2}} (-\Delta_L(\x, \z_{2}))
=-\Delta_L(\z_{1}, \z_{2})=1,
$$
by (\ref{eqn:Delta}).
Then if $h(\z_{1})<1$ and $h(\z_{2}) < 1$,
the condition of FSC (\ref{eqn:FSC}) is satisfied. \qed
\vskip 0.3cm
By Propositions \ref{thm:stationary} and \ref{thm:RandA_2},
the above lemma implies the following.
\begin{prop}
\label{thm:P00_1}
For any $L \geq 2$,
$$
P_{\alpha \beta,L}(0, \pm \e_i)=0, \quad
1 \leq i \leq d, 
\quad \alpha, \beta \in \left\{0, \frac{1}{n}, \frac{2}{n}, \dots, 1-\frac{1}{n} \right\}.
$$
Then, 
$$
P_{\alpha \beta}(0, \pm \e_i)= \lim_{L \uparrow \infty} P_{\alpha \beta, L}(0, \pm \e_i)=0, \quad
1 \leq i \leq d, 
\quad
\alpha, \beta \in \left\{0, \frac{1}{n}, \frac{2}{n}, \dots, 1-\frac{1}{n} \right\}.
$$
\end{prop}

\subsection{Determinatal expressions of $P_{0,L}(0)$
and $P_{00,L}(0, \x)$ \label{sec:matrix}}

Let $\e_{i}, 1 \leq i \leq d$ be the $i$-th unit vector in $\Z^{d}$.
Define a real symmetric matrix
with size $(2L+1)^d$ as
\begin{eqnarray}
B_L^{(0)}(\v,\w)=
\left\{
   \begin{array}{rl}
-h_{\rm c}+1/n,
 & \quad \hbox{if} \quad \v=\w=0, \\
-1, & 
\quad \hbox{if} \quad \v=\w,|\v|=1, \v \not=-\e_{d}, \\
-1+1/n, & \quad \hbox{if} \quad \v=\w=-\e_{d}, \\
1, & \quad \hbox{if} \quad \v=0, |\w|=1, \w \not=-\e_{d}, \\
1-1/n, & \quad \hbox{if} \quad \v=0, \w=-\e_{d}, \\
0, & \quad \hbox{otherwise,} \\
\end{array}\right. \nonumber
\label{eqn:B0}
\end{eqnarray}
where $\v, \w \in \Lambda_L$.

\begin{lem}
\label{thm:P0L}
Let $E_L$ be the unit matrix with size $(2L+1)^d$. Then
$$
  P_{0,L}(0)=\det \left(E_L+n G_L B_L^{(0)} \right).
$$
\end{lem}
\noindent{\it Proof.} 
Define a set of allowed configurations conditioned $h(0) =0$,
$$
\cA_L^{(0)}=
\{ h \in \cA_L: h(0) =0 \}.
$$
By definition (\ref{eqn:correlation1}), 
Proposition \ref{thm:stationary} with Lemma \ref{thm:det_L}
and Proposition \ref{thm:RandA_2} gives
\begin{equation}
P_{0,L}(0)=\frac{|\cA_L^{(0)}|}{n^{(2L+1)^d} \det \Delta_{L}}.
\label{eqn:P0L2}
\end{equation}
Assume that $h \in \cA_L^{(0)}$.
Then as shown in the proof of Lemma \ref{thm:AandT}
we can uniquely define a burning process $T_{t}, t \in \{0,1, \dots, ^{\exists} \sigma\}$
on $(G_L, h)$ associated that
$T_{t}$ becomes a spanning tree on $G_L$ at 
time $t=\sigma$. Define a configuration $h^{\prime}$ as
\begin{eqnarray}
h^{\prime}(\z)= 
\left\{
   \begin{array}{ll}
 h(\z)-1,
 & \quad \hbox{if} \quad |\z|=1, \z \not= -\e_{d}, \\
 h(\z)-1+1/n, & 
\quad \hbox{if} \quad \z=-\e_{d}, \\
 h(\z), & \quad \hbox{otherwise} \\
   \end{array}\right. \nonumber
\label{eqn:etap}
\end{eqnarray}
for $\z \in \Lambda_L$.
Now we consider a new DASM which is defined by the
matrix $\Delta_L^{\prime}$ given by
\begin{equation}
  \Delta_L^{\prime}=\Delta_L+B_L^{(0)},
\label{eqn:Delp}
\end{equation}
and let $\cA_L^{\prime}$ be a set of all allowed configurations
of this DASM and
$G_L^{\prime}$ be an associated graph to
$(\Lambda_L, \Delta_L^{\prime})$.
Then we consider a burning process 
$T_{t}^{\prime}=(V_t^{\prime}, E_t^{\prime}), t \in \{0,1,\dots, \sigma\}$
on $(G_L^{\prime}, h^{\prime})$.
By definition of $\Delta_L^{\prime}$
and $h^{\prime}$, we can make
$$
  V_{t}=V_{t}^{\prime}, \quad \forall t \in \{0, 1, \dots, \sigma \}, 
$$
and $T_{\sigma}^{\prime}$ gives a spanning tree on $G_L^{\prime}$.
By Lemma \ref{thm:AandT},
this means $h^{\prime} \in \cA_L^{\prime}$.
Since there is a bijection between $h$ and its associated
burning process $T_{t}, t \in \{0,1,\dots, \sigma\}$, we have a bijection between
$\cA_L^{(0)}$ and $\cA_L^{\prime}$.
By Lemmas \ref{thm:spanning_tree2} and \ref{thm:AandT},
$|\cA_L^{(0)}| = |\cA_L^{\prime}|
=n^{(2L+1)^d} \det \Delta_L^{\prime}$.
Combining (\ref{eqn:P0L2}) and (\ref{eqn:Delp})
gives
\begin{eqnarray}
  P_{0,L}(0) &=& \frac{\det \Delta_L^{\prime}}{\det \Delta_L} \nonumber\\
  &=& \det (\Delta_L^{-1} \Delta_L^{\prime}) \nonumber\\
  &=& \det(E_L+\Delta_L^{-1} B_L^{(0)} ).
\nonumber
\end{eqnarray}
Then we use Lemma \ref{thm:GL} and the proof is completed.
\qed
\vskip 0.3cm

\begin{figure}[htbp]
\hskip 5.5cm
\includegraphics[width=0.3\linewidth]{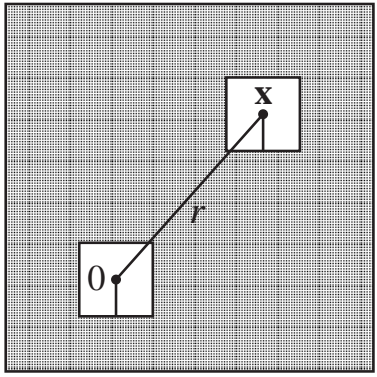}
\caption{
The matrix $\Delta_L^{\prime \prime} \equiv \Delta_L+B_L^{(0, \x)}$
is considered for $P_{00,L}(0, \x)$
with $|\x|=r$.
In the corresponding graph $G_L^{\prime \prime}$
the site 0 (resp. $\x$) is connected to
$-\e_d$ (resp. $\x-\e_d$) by a single edge,
but all other edges between 0 (resp. $\x$)
and its nearest-neighbor sites are deleted.
}
\label{fig:graph_G''}
\end{figure}

Next we consider the two-point function
$P_{00,L}(0, \x)$, where we assume that $2 \leq |\x| < L$.
We define a real symmetric matrix with size $(2L+1)^d$ as follows.
For $\v, \w \in \Lambda_L$, 
\begin{eqnarray}
B_{L}^{(0, \x)}(\v, \w)=
\left\{
   \begin{array}{rl}
-h_{\rm c}+1/n,
 & \quad \hbox{if} \quad \v=\w=0 \ \hbox{or if} \ \v=\w=\x, \\
-1, & 
\quad \hbox{if} \quad \v=\w,|\v|=1, \v \not=-\e_{d}, \\
 & 
\quad \hbox{or if} \quad \v=\w,|\v-\x|=1, \v \not=\x-\e_{d}, \\
-1+1/n, & \quad \hbox{if} \quad \v=\w=-\e_{d}, \quad
 \hbox{or if} \quad
\v=\w=\x-\e_{d}, \\
1, & \quad \hbox{if} \quad \v=0, |\w|=1, \w \not=-\e_{d}, \\
  & \quad \hbox{or if} \quad \v=\x, |\w-\x|=1, \w \not=\x-\e_{d} \\
1-1/n, & \quad \hbox{if} \quad \v=0, \w=-\e_{d}, \\
    & \quad \hbox{or if} \quad \v=\x, \w=\x-\e_{d}, \\
0, & \quad \hbox{otherwise.} \\
\end{array}\right. \nonumber
\label{eqn:B0x}
\end{eqnarray}
Following the same argument as $P_{0,L}(0)$ we can prove
the next lemma. (See Fig.\ref{fig:graph_G''}.) 

\begin{lem}
\label{thm:P00L}
For $2 \leq |\x| < L$, 
$$
  P_{00,L}(0,\x) = \det
  \left( E_L+n G_L B_L^{(0, \x)} \right).
$$
\end{lem}

\subsection{Infinite-volume limit \label{sec:infinite}}

Since the number of nonzero elements of $B_L^{(0)}$
(resp. $B_L^{(0,\x)}$ ) is only $6d+1$ (resp. $2(6d+1)$),
we can replace the matrix $E_L+nG_L B_L^{(0)}$ 
(resp. $E_L+nG_L B_L^{(0,\x)}$) with size $(2L+1)^d$
by a matrix with size $(2d+1)$
(resp. $2(2d+1)$) without changing the value of determinant.
Explicit expressions are given as follows.

Let
$$
\q_{i}=
\left\{
   \begin{array}{ll}
      0, & \quad \hbox{if} \quad i=1, \\
      \e_{i-1}, & \quad \hbox{if} \quad 2 \leq i \leq d+1, \\
      -\e_{i-d-1}, & \quad \hbox{if} \quad d+2 \leq i \leq 2d+1. \\
   \end{array}\right.
$$
Define a matrix
$\cG^{(L)}(\x)=(\cG^{(L)}_{ij})_{1 \leq i, j \leq 2d+1}$ with elements
\begin{equation}
  \cG^{(L)}_{ij}(\x)
  =G_{L}(0, \x+\q_{j}-\q_{i}), \quad 1 \leq i, j \leq 2d+1.
\label{eqn:GLij}
\end{equation}
We also define a real symmetric
matrix $\cB=(\cB_{ij})_{1 \leq i, j \leq 2d+1}$ with elements
\begin{eqnarray}
\cB_{ij}=
\left\{
   \begin{array}{ll}
 -h_{\rm c}+1/n,
 & \quad \hbox{if} \quad i=j=1, \\
  -1, & \quad \hbox{if} \quad 2 \leq i=j \leq 2d, \\
  -1+1/n, & \quad \hbox{if} \quad i=j=2d+1, \\
  1, & \quad i=1, 2 \leq j \leq 2d, \\
  1-1/n, & \quad \hbox{if} \quad i=1, j=2d+1, \\
  0, & \quad \hbox{otherwise.} \\
   \end{array}\right. \nonumber
\end{eqnarray}
Then define $2(2d+1) \times 2(2d+1)$ matrices
$$
  \tilde{\cG}^{(L)}(0,\x)
  =\left( \begin{array}{ll} 
\cG^{(L)}(0) & \cG^{(L)}(\x) \cr
       ^{t}\cG^{(L)}(\x) & \cG^{(L)}(0) 
\end{array} \right), \quad \x \in \Lambda_L, 
$$
where $^{t}\cG^{(L)}(\x)$ is a transpose of
$\cG^{(L)}(\x)$,
and 
$$
   \tilde{\cB}=\left( 
\begin{array}{ll} 
\cB & 0 \cr
       0 & \cB 
\end{array} \right).
$$
We have
\begin{equation}
  P_{0,L}(0)=\det \left(E+ n \cG^{(L)}(0)
  \cB \right)
\label{eqn:P0L0}
\end{equation}
and
\begin{equation}
  P_{00,L}(0,\x)=\det \left(E + n
  \tilde\cG^{(L)}(0,\x) \tilde\cB \right),
\label{eqn:P00L0}
\end{equation}
where $E$ denotes the unit matrix
with size $2d+1$ in (\ref{eqn:P0L0})
and with size $2(2d+1)$ in (\ref{eqn:P00L0}), 
respectively.

It should be remarked that the sizes
of the matrices in the RHS's are independent
of the lattice size $L$ and determined only by
the dimension $d$ of lattice.
The dependence of $L$ is introduced only
through each elements of 
$\cG^{(L)}(\x)$ given by (\ref{eqn:GLij}).
Lemma \ref{thm:G_infinite} guarantees the existence
of infinite-volume limit $L \uparrow \infty$
of these elements and we put
\begin{eqnarray}
\cG_{ij}(\x) &=& \lim_{L \uparrow \infty}  \cG^{(L)}_{ij}(\x) 
=G(\x+\q_{j}-\q_{i}), \quad 1 \leq i, j \leq 2d+1,
\nonumber\\
\cG(\x) &=& (\cG_{ij}(\x))_{1 \leq i, j \leq 2d+1},
\nonumber\\
\tilde{\cG}(0,\x)
&=& \lim_{L \uparrow \infty} \tilde{\cG}^{(L)}(0,\x)
=\left( \begin{array}{ll} \cG(0) & \cG(\x) \cr
       ^{t}\cG(\x) & \cG(0) 
\end{array} \right),
\nonumber
\end{eqnarray}
where $G(\x)$ is explicitly given by
(\ref{eqn:Green4}).
Then we have the following.
\begin{prop}
\label{thm:infinite_P}
There exist the infinite-volume limits
$$
P_{0}=\lim_{L \uparrow \infty}P_{0, L}(0),
\quad
P_{00}(\x)=\lim_{L \uparrow \infty} P_{00,L}(0,\x), \quad \x \in \Z^d, 
$$ 
and they are given by
$$
  P_{0}=\det \left(E+ n  \cG(0)
  \cB \right)
$$
and
$$
  P_{00}(\x)=\det \left(E + n
  \tilde\cG(0,\x) \tilde\cB \right), \quad \x \in \Z^d.
$$
\end{prop}

\subsection{Evaluations of determinantal expressions \label{sec:asym}}

From the determinantal expressions of $P_0$ and $P_{00}(\x)$
given in Proposition \ref{thm:infinite_P}, 
the following explicit evaluations of these quantities are obtained.

\begin{thm}
\label{thm:P0andP00}
\begin{description}
\item{{\rm (i)}} \quad
Define
$$
  \gamma_{1}
  = \frac{1}{2d} \prod_{i=1}^d \int_{-\pi}^{\pi} \frac{d \theta_{i}}{2 \pi}
\frac{1}
{(1+a) - (1/d) \sum_{i=1}^{d} \cos \theta_{i}}
$$
and
$$
  \gamma_{2}
  = \frac{1}{2d} \prod_{i=1}^d \int_{-\pi}^{\pi} \frac{d \theta_{i}}{2 \pi}
\frac{ {\rm e}^{-2 \sqrt{-1} (\theta_{1}+\theta_{2})} }
{(1+a) - (1/d) \sum_{i=1}^{d} \cos \theta_{i}}.
$$
Then, for the DASM with $d \geq 2, m, n \in \N$, 
\begin{eqnarray}
P_{0} &=& 
\frac{1-2da \gamma_{1}}{2dn}
\left[ 2 \{1-d(\gamma_{1}-\gamma_{2})\}
+(1-4 d \gamma_{1}) a 
- 2 d \gamma_{1} a^{2} \right]
\nonumber\\
&\times& \left[2(d-1) (\gamma_{1}-\gamma_{2})
-(1-4 d \gamma_{1})a 
+ 2 d \gamma_{1} a^{2} \right]^{2}
\nonumber\\
&\times& \left[ 
\{1-(\gamma_{1}-\gamma_{2}) \}^{2}
-\{(2d(1+a)^{2}-1)
\gamma_{1}-(2d-1)\gamma_{2}
-(1+a) \}^{2} \right]^{d-2},
\label{eqn:P0}
\end{eqnarray}
where $a=m/(2dn)$.
\item{{\rm (ii)}} \quad
Let
\begin{equation}
  C_{00}(\x)=\frac{P_{00}(\x)-P_{0}^{2}}{P_{0}^{2}}, 
\quad \x \in \Z^d.
\label{eqn:C002}
\end{equation}
Then, there exists a nonzero factor 
$c_{2}(d,a,n)$ such that
for the DASM with $d \geq 2, m, n \in \N$
\begin{equation}
\lim_{r \uparrow \infty} - \frac{1}{r}
\log \left[ \frac{r^{d-1}}{c_{2}(d,a,n)}
C_{00}(\x(r)) \right]
= \frac{2}{\xi(d, a)},
\label{eqn:thm21}
\end{equation}
where 
$a=m/(2dn)$, $\xi(d, a)$ and $\x(r)$ are given by (\ref{eqn:xi}) and (\ref{eqn:xr}),
respectively, 
and that
\begin{equation}
\lim_{a \downarrow 0} 
\frac{c_{2}(d, a, m/(2da))}
{a^{(d+1)/2}}
= \left(\frac{d}{2 \pi^{2}} \right)^{(d-3)/2}
\left[ \frac{d \{1+(d-1) \bar{\gamma} \}}
{2 \pi (d-1) \bar{\gamma} } \right]^{2},
\label{eqn:limA}
\end{equation}
where
$$
  \bar{\gamma}=
\frac{1}{2d} \prod_{i=1}^d \int_{-\pi}^{\pi} \frac{d \theta_{i}}{2 \pi}
\frac{ 1-{\rm e}^{-2 \sqrt{-1} (\theta_{1}+\theta_{2})} }
{1- (1/d) \sum_{i=1}^{d} \cos \theta_{i}}.
$$
\end{description}
\end{thm}
\vskip 0.3cm

In the following, we will explain how to prove this theorem.
Let
$$
  M^{(1)}(r)=E + n \tilde{\cG}(0,\x(r))
  \tilde{\cB}, \quad r >0, \quad \x(r) \in \Z^d, 
$$
where $E$ is a unit matrix with size $2(2d+1)$.
That is,
$$
   M^{(1)}(r)=\left( \begin{array}{ll} m^{(1)} & \tilde{m}^{(1)}(r) \cr
       \hat{m}^{(1)}(r) & m^{(1)} 
\end{array} \right),
$$
where for $1 \leq i \leq 2d+1$
\begin{eqnarray}
m^{(1)}_{ij}= 
\left\{
   \begin{array}{ll}
\1(i=1)+ \sum_{k=1}^{2d+1} n\cG_{ik}(0)
& \\ \quad 
-\{(1-1/n)+h_{\rm c} \}n\cG_{i1}(0)
-\cG_{i \, 2d+1}(0), 
 & \quad \hbox{if} \quad j=1, \\
 \1(i=j)+n [\cG_{i 1}(0)-\cG_{i j}(0) ], & 
\quad \hbox{if} \quad 2 \leq j \leq 2d, \\
\1(i=2d+1) +
(1-1/n)n[\cG_{i 1}(0)-\cG_{i \, 2d+1}(0)],
& \quad \hbox{if} \quad j=2d+1,\\
   \end{array}\right. \nonumber
\end{eqnarray}
\begin{eqnarray}
\tilde{m}^{(1)}_{ij}(r)=n \times
\left\{
   \begin{array}{ll}
\sum_{k=1}^{2d+1} \cG_{i k}(\x(r))
& \\ \quad 
-\{(1-1/n)+h_{\rm c} \}\cG_{i 1}(\x(r))
-(1/n) \cG_{i \, 2d+1}(\x(r)), 
 & \quad \hbox{if} \quad j=1, \\
\cG_{i 1}(\x(r))-\cG_{i j}(\x(r)), & 
\quad \hbox{if} \quad 2 \leq j \leq 2d, \\
(1-1/n)(\cG_{i 1}(\x(r))-\cG_{i \, 2d+1}(\x(r))), & 
\quad \hbox{if} \quad 
j=2d+1,\\
   \end{array}\right. \nonumber
\end{eqnarray}
\begin{eqnarray}
\hat{m}^{(1)}_{ij}(r)=n \times
\left\{
   \begin{array}{ll}
\sum_{k=1}^{2d+1} \cG_{k i}(\x(r))
& \\ \quad 
-\{(1-1/n)+ \eta_{{\rm c}} \} \cG_{1 i}(\x(r))
-(1/n) \cG_{2d+1 \, i}(\x(r)), 
 & \quad \hbox{if} \quad j=1, \\
\cG_{1 i}(\x(r))-\cG_{j i}(\x(r)), & 
\quad \hbox{if} \quad 2 \leq j \leq 2d, \\
(1-1/n)(\cG_{1 i}(\x(r))-\cG_{2d+1 \, i}(\x(r))), & 
\quad \hbox{if} \quad 
j=2d+1.\\
   \end{array}\right. \nonumber
\end{eqnarray}
We find that
\begin{eqnarray}
&& m^{(1)}_{i1}+\sum_{j=2}^{2d+1} m^{(1)}_{ij}
=1-2dan \cG_{i 1}(0), \nonumber\\
&& \tilde{m}^{(1)}_{i1}(r)+\sum_{j=2}^{2d+1} \tilde{m}^{(1)}_{ij}
=-2dan \cG_{i 1}(\x(r)), \nonumber\\
&& \hat{m}^{(1)}_{i1}(r)+\sum_{j=2}^{2d+1} \hat{m}^{(1)}_{ij}
=-2dan \cG_{1 i}(\x(r)), \quad 1 \leq i \leq 2d+1.
\nonumber
\end{eqnarray}
For $1 \leq i \leq 2d+1$, let
\begin{eqnarray}
m_{ij} &=&
\left\{
   \begin{array}{ll}
1-2dan \cG_{i 1}(0),
 & \quad \hbox{if} \quad j=1, \\
m^{(1)}_{ij}, & 
\quad \hbox{if} \quad 2 \leq j \leq 2d+1, \\
   \end{array}\right. \nonumber\\
\tilde{m}_{ij}(r) &=&
\left\{
   \begin{array}{ll}
-2dan \cG_{i 1}(\x(r)), 
 & \quad \hbox{if} \quad j=1, \\
\tilde{m}^{(1)}_{ij}(r), & 
\quad \hbox{if} \quad 2 \leq j \leq 2d+1, \\
   \end{array}\right. \nonumber\\
\hat{m}_{ij}(r) &=&
\left\{
   \begin{array}{ll}
-2dan \cG_{1 i}(\x(r)),
 & \quad \hbox{if} \quad j=1, \\
\hat{m}^{(1)}_{ij}(r), & 
\quad \hbox{if} \quad 2 \leq j \leq 2d+1. \\
   \end{array}\right.  \nonumber
\end{eqnarray}
Then
\begin{eqnarray}
P_0 &=& \det m^{(1)} = \det m,
\nonumber\\
P_{00}(\x(r)) &=& 
 \det M^{(1)}(r) = \det M(r) \quad \hbox{with} \quad
   M(r)=\left( \begin{array}{ll} m & \tilde{m}(r) \cr
       \hat{m}(r) & m 
\end{array} \right).
\label{eqn:det_exp_A1}
\end{eqnarray}
Note that, if we introduce the 
the {\it dipole potential}
$$
\phi_{(i_{1}, j_{1}), (i_{2}, j_{2})}(\x(r))
=\cG_{i_{1} j_{1}}(\x(r))-
\cG_{i_{2} j_{2}}(\x(r)), 
\quad 1 \leq i_1, i_2, j_1, j_2 \leq 2d+1, 
$$
the elements of the matrix $M(r)$ are expressed as follows;
for $1 \leq i \leq 2d+1$, 
\begin{eqnarray}
m_{ij}=
\left\{
   \begin{array}{ll}
1-2dan \cG_{i1}(0),
 & \quad \hbox{if} \quad j=1, \\
 \1(i=j)+n \phi_{(i,1),(i,j)}(0), & 
\quad \hbox{if} \quad 2 \leq j \leq 2d, \\
\1(i=2d+1)+
(1-1/n) n \phi_{(i,1),(i,2d+1)}(0),
& \quad \hbox{if} \quad j=2d+1,\\
   \end{array}\right.
\label{eqn:m}
\end{eqnarray}
\begin{eqnarray}
\tilde{m}_{ij}(r)= n \times
\left\{
   \begin{array}{ll}
-2da \cG_{i 1}(\x(r)),
 & \quad \hbox{if} \quad j=1, \\
\phi_{(i,1),(i,j)}(\x(r)), & 
\quad \hbox{if} \quad 2 \leq j \leq 2d, \\
(1-1/n)  \phi_{(i,1),(i,2d+1)}(\x(r)), & 
\quad \hbox{if} \quad 
j=2d+1,\\
   \end{array}\right. \nonumber
\end{eqnarray}
\begin{eqnarray}
\hat{m}_{ij}(r)= n \times
\left\{
   \begin{array}{ll}
-2da  \cG_{1i}(\x(r)),
 & \quad \hbox{if} \quad j=1, \\
\phi_{(1,i),(j,i)}(\x(r)), & 
\quad \hbox{if} \quad 2 \leq j \leq 2d, \\
(1-1/n) \phi_{(1,i),(2d+1,i)}(\x(r)), & 
\quad \hbox{if} \quad 
j=2d+1.\\
   \end{array}\right. \nonumber
\end{eqnarray}

Now we study the asymptotics of $P_{00}(r)$ in $r \uparrow \infty$.
Theorem \ref{thm:main1} and its proof given in Section \ref{sec:propagator}
implies that with any finite $c_{i}$'s, 
$$
  G \left(\x(r) + \sum_{i=1}^{d} c_{i} \e_{i} \right)
= \bar{G}(r) \exp \left( - \lambda(a) \sum_{i=1}^{d} c_{i} \right)
\times \left( 1 + \cO(1/r) \right), 
\quad \mbox{as $r \uparrow \infty$}
$$
with (\ref{eqn:c1}),(\ref{eqn:xi}), (\ref{eqn:lambda}), and (\ref{eqn:Gbar}). 
Then we see 
\begin{eqnarray}
\tilde{m}(r) &=& n \bar{G}(r) n(r, \lambda) (1+\cO(1/r)), \nonumber\\
\hat{m}(r) &=& n \bar{G}(r) n(r, -\lambda) (1+\cO(1/r)), \quad
\mbox{as $r \uparrow \infty$}, 
\nonumber
\end{eqnarray}
where $n(r, \lambda)=(n_{ij}(r, \lambda))_{1 \leq i, j \leq 2d+1}$
with elements,
\begin{equation}
n_{ij}(r, \lambda) 
=
\left\{
   \begin{array}{ll}
     -2da, & \quad \hbox{if} \quad i=j=1, \\
      (1-{\rm e}^{-\lambda}),
& \quad \hbox{if} \quad i=1, 2 \leq j \leq d+1, \\
      (1-{\rm e}^{\lambda}),
& \quad \hbox{if} \quad i=1, d+2 \leq j \leq 2d, \\
      (1-1/n) (1-{\rm e}^{\lambda}),
& \quad \hbox{if} \quad i=1, j=2d+1, \\
     -2da  {\rm e}^{\lambda},
   & \quad \hbox{if} \quad 2 \leq i \leq d+1, j=1, \\
     -2da {\rm e}^{-\lambda},
   & \quad \hbox{if} \quad d+2 \leq i \leq 2d+1, j=1, \\
   {\rm e}^{\lambda}(1-{\rm e}^{-\lambda}),
   & \quad \hbox{if} \quad 2 \leq i, j \leq d+1, \\
   {\rm e}^{\lambda}(1-{\rm e}^{\lambda}),
   & \quad \hbox{if} \quad 2 \leq i \leq d+1, d+2 \leq j \leq 2d, \\
  (1-1/n) {\rm e}^{\lambda}(1-{\rm e}^{\lambda}), 
   & \quad \hbox{if} \quad 2 \leq i \leq d+1, j=2d+1, \\
   {\rm e}^{-\lambda}(1-{\rm e}^{-\lambda}),
   & \quad \hbox{if} \quad d+2 \leq i \leq 2d+1, 2 \leq j \leq d+1, \\
   {\rm e}^{-\lambda}(1-{\rm e}^{\lambda}),
   & \quad \hbox{if} \quad d+2 \leq i, j \leq 2d, \\
   (1-1/n){\rm e}^{-\lambda}(1-{\rm e}^{\lambda}),
   & \quad \hbox{if} \quad d+2 \leq i \leq 2d+1, j=2d+1. \\
   \end{array}\right.  
\nonumber
\end{equation}

We obtain a matrix $M^{\prime}(r)$ from $M(r)$
by subtracting (the first row) $\times {\rm e}^{\lambda}$
from the $i$-th row with $2 \leq i \leq d+1$,
(the first row) $\times {\rm e}^{-\lambda}$
from the $i$-th row with $d+2 \leq i \leq 2d+1$,
(the $(2d+2)$-th row) $\times {\rm e}^{-\lambda}$
from the $i$-th row with $2d+3 \leq i \leq 3d+2$,
and
(the $(2d+2)$-th row) $\times {\rm e}^{\lambda}$
from the $i$-th row with $3d+3 \leq i \leq 2(2d+1)$.
We have
$$
   M^{\prime}(r)=\left( 
\begin{array}{ll} m^{\prime}(\lambda)
    & \tilde{m}^{\prime}(r, \lambda) \cr
       \tilde{m}^{\prime}(r, -\lambda) 
       & m^{\prime}(-\lambda) \cr
\end{array} \right)
$$
with
\begin{eqnarray}
m^{\prime}_{ij}(\lambda) &=&
\left\{
   \begin{array}{ll} 
      1-2da n \cG_{11}(0), & \hbox{if} \quad i=j=1, \\
      n \phi_{(1,1),(1,j)}(0),
   & \hbox{if} \quad i=1, 2 \leq j \leq 2d, \\
      (1-1/n) n \phi_{(1,1),(1,2d+1)}(0),
   & \hbox{if} \quad i=1, j=2d+1, \\
      (1-{\rm e}^{\lambda})
       -2da n (\cG_{i1}(0)-{\rm e}^{\lambda}\cG_{11}(0)),
     & \hbox{if} \quad 2 \leq i \leq d+1, j=1, \\
       (1-{\rm e}^{-\lambda})
       -2da n (\cG_{i1}(0)-{\rm e}^{-\lambda}\cG_{11}(0)),
     & \hbox{if} \quad d+2 \leq i \leq 2d+1, j=1, \\
    \1(i=j) +n[\phi_{(i,1),(i,j)}(0)
       -{\rm e}^{\lambda} \phi_{(1,1),(1,j)}(0)],
     & \hbox{if} \quad
    2 \leq i \leq d+1, 2 \leq j \leq 2d, \\
    \1(i=j)+n[ \phi_{(i,1),(i,j)}(0)
       -{\rm e}^{-\lambda} \phi_{(1,1),(1,j)}(0)],
     & \hbox{if} \quad
    d+2 \leq i \leq 2d+1, 2 \leq j \leq 2d, \\
   (1-1/n) & \\
   \quad \times n [\phi_{(i,1),(i,2d+1)}(0)
     -{\rm e}^{\lambda} \phi_{(1,1),(1,2d+1)}(0) ],
     & \hbox{if} \quad
    2 \leq i \leq d+1, j=2d+1, \\
   \1(i=2d+1)+(1-1/n) & \\
   \quad \times n [\phi_{(i,1),(i,2d+1)}(0)
     -{\rm e}^{-\lambda} \phi_{(1,1),(1,2d+1)}(0)],
     & \hbox{if} \quad
    d+2 \leq i \leq 2d+1, j=2d+1, \\
   \end{array}\right. 
\nonumber\\
\label{eqn:mprime}
\end{eqnarray}
and with
\begin{eqnarray}
\tilde{m}^{\prime}_{ij}(r, \lambda)=
n\bar{G}(r) \times
\left\{
   \begin{array}{ll}
     -2da(1+\cO(1/r)),  & \hbox{if} \quad i=j=1, \\
      (1-{\rm e}^{-\lambda})(1+\cO(1/r)),
& \hbox{if} \quad i=1, 2 \leq j \leq d+1, \\
       (1-{\rm e}^{\lambda}) (1+\cO(1/r)),
& \hbox{if} \quad i=1, d+2 \leq j \leq 2d, \\
       (1-1/n) (1-{\rm e}^{\lambda})(1+\cO(1/r)),
& \hbox{if} \quad i=1, j=2d+1, \\
\cO(1/r),
& \quad \hbox{otherwise},\\ 
   \end{array}\right. 
\label{eqn:tilmp}
\end{eqnarray}
so that
$$
 P_00(\x(r))=\det M(r)=\det M^{\prime}(r), \quad r >0, \quad \x(r) \in \Z^d.
$$
Now we expand $\det M^{\prime}(r)$ along the first and the $(2d+2)$-th
rows. Let $|M^{\prime}(j,k)|$ be the determinant of
$M^{\prime}(r)$ with the first and the $(2d+2)$-th rows and the
$j$-th and the $k$-th columns removed and multiplied by
$-(-1)^{1+j} \times (-1)^{2d+2+k}=(-1)^{j+k}$.
Then we have
$$
  \det M^{\prime}(r)=\sum_{j=1}^{2(2d+1)} 
\sum_{k=1, k \not= j}^{2(2d+1)}
M^{\prime}(r)_{1j} M^{\prime}(r)_{2d+2,k}
|M^{\prime}(j,k)|.
$$
Remark that, by (\ref{eqn:mprime}) and (\ref{eqn:tilmp}),
$$
|M^{\prime}(j,k)|=\cO(1/r), \quad \mbox{as $r \to \infty$}, 
$$
if
$ 1 \leq j, k \leq 2d+1$ or
$2d+2 \leq j, k \leq 2(2d+1)$,
and
$$
 |M^{\prime}(j,k)|=|m^{\prime (j)}(\lambda)| 
\times |m^{\prime (k)}(\lambda)|
\times (1+\cO(1/r)), \quad \mbox{as $r \to \infty$}, 
$$
if
$1 \leq j \leq 2d+1 < k \leq 2(2d+1)$
or
$1 \leq k \leq 2d+1 < j \leq 2(2d+1)$,
where $|m^{\prime (j)}(\lambda)|$ is the 
$(1,j)$-cofactor of $m^{\prime}(\lambda)$.
Then
\begin{eqnarray}
&& \det M^{\prime}(r)
=
\left( \sum_{j=1}^{2d+1} m^{\prime}_{1j}(\lambda)
 |m^{\prime (j)}(\lambda)| \right)
\left( \sum_{j=1}^{2d+1} m^{\prime}_{1j}(-\lambda)
 |m^{\prime (j)}(-\lambda)| \right) \nonumber\\
&& \qquad \qquad +
\left( \sum_{j=1}^{2d+1} \tilde{m}^{\prime}_{1j}(r, \lambda)
 |m^{\prime (j)}(-\lambda)| \right)
\left( \sum_{j=1}^{2d+1} \tilde{m}^{\prime}_{1j}(r, -\lambda)
 |m^{\prime (j)}(-\lambda)| \right)
\nonumber\\
&& \quad = \det m^{\prime}(\lambda) \times \det m^{\prime}(-\lambda)
+\det \bar{m}(\lambda) \times \det \bar{m}(-\lambda) \times 
\left(n \bar{G}(r) \right)^2
(1+\cO(1/r)), 
\label{eqn:detMp}
\end{eqnarray}
where $\bar{m}(\lambda)=(\bar{m}_{ij}(\lambda))_{1 \leq i, j \leq 2d+1}$ with elements
\begin{eqnarray}
\bar{m}_{ij}(\lambda)=
\left\{
   \begin{array}{ll}
     -2da, & \quad \hbox{if} \quad i=j=1, \\
      1-{\rm e}^{\lambda},
& \quad \hbox{if} \quad i=1, 2 \leq j \leq d+1, \\
      1-{\rm e}^{-\lambda},
& \quad \hbox{if} \quad i=1, d+2 \leq j \leq 2d, \\
       (1-1/n) (1-{\rm e}^{-\lambda}),
& \quad \hbox{if} \quad i=1, j=2d+1, \\
 m^{\prime}_{ij}(\lambda),
& \quad \hbox{otherwise}.\\ 
   \end{array}\right.  \nonumber
\label{eqn:mbar}
\end{eqnarray}
We find that 
\begin{equation}
\det m^{\prime}(\lambda)
=\det m^{\prime}(-\lambda)= \det m.
\label{eqn:detm_eq}
\end{equation}
The determinantal expressions (\ref{eqn:det_exp_A1}) with
(\ref{eqn:Gbar}), (\ref{eqn:detMp}), and (\ref{eqn:detm_eq}) give
\begin{eqnarray}
\lim_{r \uparrow \infty} P_{00}(\x(r))
&=& \lim_{r \uparrow \infty}
\{ (\det m)^2+\det \bar{m}(\lambda) \det \tilde{m}(-\lambda) (n\bar{G}(r))^2 \}
\nonumber\\
&=& (\det m)^2 = P_0^2.
\nonumber
\end{eqnarray}

Here we set
$$
  \det \bar{m}(\lambda)=a \det m^{*}(\lambda),
$$
with a matrix 
$m^{*}(\lambda)=(m^*_{ij}(\lambda))_{1 \leq i, j \leq 2d+1}$ with elements
\begin{eqnarray}
m^{*}_{ij}(\lambda)= 
\left\{
   \begin{array}{ll}
      -2d, & \hbox{if} \, \, i=j=1, \\
      (1-{\rm e}^{\lambda})/a^{1/2},
   & \hbox{if} \, \, i=1, 2 \leq j \leq d+1, \\
     (1-{\rm e}^{-\lambda})/a^{1/2},
   & \hbox{if} \, \, i=1, d+2 \leq j \leq 2d, \\
      (1-1/n) (1-{\rm e}^{-\lambda})/a^{1/2},
   & \hbox{if} \, \, i=1, j=2d+1, \\
      (1-{\rm e}^{\lambda})/a^{1/2} & \\
   \quad -2da^{1/2}n (\cG_{i1}(0)
-{\rm e}^{\lambda}\cG_{11}(0)),
     & \hbox{if} \, \, 2 \leq i \leq d+1, j=1, \\
       (1-{\rm e}^{-\lambda})/a^{1/2} & \\
   \quad -2da^{1/2} n (\cG_{i1}(0)
-{\rm e}^{-\lambda} \cG_{11}(0)),
     & \hbox{if} \, \, d+2 \leq i \leq 2d+1, j=1, \\
   \1(i=j)+n[ \phi_{(i,1),(i,j)}(0)
       -{\rm e}^{\lambda} \phi_{(1,1),(1,j)}(0)],
     & \hbox{if} \,\, 
    2 \leq i \leq d+1, 2 \leq j \leq 2d, \\
   \1(i=j)+n[\phi_{(i,1),(i,j)}(0)
       -{\rm e}^{-\lambda} \phi_{(1,1),(1,j)}(0)],
     & \hbox{if} \,\, 
    d+2 \leq i \leq 2d+1, 2 \leq j \leq 2d, \\
   (1-1/n) & \\
   \ \times n [\phi_{(i,1),(i,2d+1)}(0)
     -{\rm e}^{\lambda} \phi_{(1,1),(1,2d+1)}(0) ],
     & \hbox{if} \,\, 
    2 \leq i \leq d+1, j=2d+1, \\
  \1(i=2d+1)+(1-1/n), & \\
   \ \times n [\phi_{(i,1),(i,2d+1)}(0)
     -{\rm e}^{-\lambda} \phi_{(1,1),(1,2d+1)}(0)],
     & \hbox{if} \,\, 
    d+2 \leq i \leq 2d+1, j=2d+1. \\
   \end{array}\right. 
\label{eqn:mst}
\end{eqnarray}
By the definition (\ref{eqn:C002}), we see
$$
  C_{00}(\x(r))= a^{2}
    \frac{\det m^{*}(\lambda) \det m^{*}(-\lambda)}
    {(\det m)^2 }
(n \bar{G}(r))^2  \times (1+\cO(1/r)),
\quad \mbox{as $r \uparrow \infty$}. 
$$
Since $\bar{G}(r)$ is given by (\ref{eqn:Gbar}),
(\ref{eqn:thm21}) of Theorem \ref{thm:P0andP00} (ii) is proved with
$$
  c_{2}(d, a, n)= (a c_{1}(d,a))^{2}
\frac{\det m^{*}(\lambda) \times \det m^{*}(-\lambda)}
{(\det m)^{2}}.
$$

Now the problem is reduced to the calculation
of $\det m$ and $\det m^*(\lambda)$.
Consider a matrix $R=(R_{ij})_{1 \leq i, j \leq N}$ with elements
\begin{eqnarray}
R_{ij}=
\left\{
   \begin{array}{ll}
 u, & \quad \hbox{if} \quad
 i=j=1,
 \\
  b, & \quad \hbox{if} \quad
 i=1, 2 \leq j \leq d+1,
 \\
  c, & \quad \hbox{if} \quad
 i=1, d+2 \leq j \leq 2d,
 \\
  (1-1/n)c, & \quad \hbox{if} \quad
 i=1, j=2d+1,
 \\
  q, & \quad \hbox{if} \quad
2 \leq i \leq d+1, j=1,
 \\
   e, & \quad \hbox{if} \quad
d+2 \leq i \leq 2d+1, j=1,
 \\
  f,  & \quad \hbox{if} \quad
2 \leq i \leq d+1, 2 \leq j \leq 2d, j \not=i,
j \not= i+d,
 \\
 1+v,   & \quad \hbox{if} \quad
2 \leq i=j \leq d+1,
 \\
  h,  & \quad \hbox{if} \quad
2 \leq i \leq d, j = i+d,
 \\
  (1-1/n)f,  & \quad \hbox{if} \quad
2 \leq i \leq d, j=2d+1,
 \\
   (1-1/n)h, & \quad \hbox{if} \quad
i=d+1, j=2d+1,
 \\
  s,  & \quad \hbox{if} \quad
d+2 \leq i \leq 2d+1, 2 \leq j \leq 2d, j \not= i,
j \not= i-d,
 \\
  t,  & \quad \hbox{if} \quad
d+2 \leq i \leq 2d+1, j=i-d,
 \\
  1+k,  & \quad \hbox{if} \quad
d+2 \leq i=j \leq 2d,
 \\
  (1-1/n)s, & \quad \hbox{if} \quad
d+2 \leq i \leq 2d, j=2d+1,
 \\
  1+(1-1/n) k,  & \quad \hbox{if} \quad
i=j=2d+1.
 \\
   \end{array}\right. 
\label{eqn:R1}
\end{eqnarray}
We perform the following procedure on $R$.
\begin{description}
\item{(i)}
Subtract (the first row) $\times q/u$
from the $i$-th row with $2 \leq i \leq d+1$.
\item{(ii)}
Subtract (the first row) $\times e/u$
from the $i$-th row with $d+2 \leq i \leq 2d+1$.
\item{(iii)}
Subtract the second row
from the $i$-th row with $3 \leq i \leq d+1$.
\item{(iv)}
Subtract the $(d+2)$-th row 
from the $i$-th row with $d+3 \leq i \leq 2d+1$.
\item{(v)}
Add the $j$-th column
to the second column with $3 \leq j \leq d+1$.
\item{(vi)}
Add the $j$-th column
to the $(d+2)$-th column with $d+3 \leq j \leq 2d$.
\item{(vii)}
Add (the $(2d+1)$-th column) $\times 1/(1-1/n)$
to the $(d+2)$-th column.
\item{(viii)}
Subtract (the $(d+j)$-th column) $\times
(t-s)/(1+k-s)$ from the $j$-th column with
$3 \leq j \leq d$.
\end{description}
After these procedures, by changing the orders of rows and
columns appropriately, we obtain the following identity.
\begin{equation}
\det R= u \times
\left[ 1+v-f-\frac{t-s}{1+k-s}(h-f) \right]^{d-2}
\times (1+k-s)^{d-2} \times \det S,
\label{eqn:detR1}
\end{equation}
where $S=(S_ij)_{1 \leq i, j \leq 4}$ with elements
$$
\begin{array}{ll} 
S_{11} = 1+v+(d-1)f-d bq/u, \quad
&S_{12} = h+(d-1)f-d cq/u, \cr
S_{13} = (1-1/n)
(f-cq/u), \quad
&S_{14} = f-bq/u, \cr
S_{21} = t+(d-1)s-d be/u, \quad
&S_{22} = 1+k+(d-1)s-d ce/u, \cr
S_{23} =(1-1/n)(s-ce/u), \quad
&S_{24} = s-be/u, \cr
S_{31} = 0, \quad
&S_{32} = 1/(n-1), \cr
S_{33} = 1+(1-1/n) (k-s), \quad
&S_{34} = t-s, \cr
S_{41} = 0, \quad
&S_{42} = 0, \cr
S_{43} = (1-1/n) (h-f), \quad
&S_{44} = 1+v-f. 
\end{array}
$$

Define 
$$
g_{0}=nG(0), \quad g_{1}=nG(\e_{1}), \quad
g_{2}=nG(2\e_{1}), \quad
g_{3}=n G(\e_{1}+\e_{2}),
$$
where $G(\x)$ is given by (\ref{eqn:Green4}) and
$\e_{1}, \e_{2}$ are the unit vectors in the first 
and second directions in $\Z^{d}$.
Since the system is isotropic, we can find that
the matrix $m$ defined by (\ref{eqn:m}) is in the form
(\ref{eqn:R1}) with
\begin{equation}
\begin{array}{ll}
u=1-2da g_{0}, \qquad
&b=c=g_{0}-g_{1}, \cr
q=e=1-2da g_{1}, \qquad
&f=s=g_{1}-g_{3}, \cr
v=k=g_{1}-g_{0}, \qquad
&h=t=g_{1}-g_{2}. 
\end{array}
\label{eqn:case1}
\end{equation}
By Lemma \ref{thm:G_infinite} and the isotropy of the
system gives
\begin{eqnarray}
&&2d(1+a) g_{0}-2dg_{1}=1, \nonumber\\
&&2d(1+a) g_{1}-(g_{0}+g_{2}+2(d-1)g_{3})=0, 
\nonumber
\end{eqnarray}
which are written as
\begin{eqnarray}
  g_{1} &=& (1+a)g_{0}-\frac{1}{2d}, \nonumber\\
  g_{2} &=& [2d(1+a)^2-1] g_{0}
  -2(d-1) g_{3}-(1+a).
\label{eqn:g12}
\end{eqnarray}
The formula (\ref{eqn:detR1}) with (\ref{eqn:case1}) and (\ref{eqn:g12}) gives
\begin{eqnarray}
P_0 = \det m &=& \frac{1-2dag_{0}}{2dn}
\left[2\{1-d(g_{0}-g_{3})\} +(1-4dg_{0})a-2dg_{0}a \right]
\nonumber\\
&\times&
\left[2(d-1)(g_{0}-g_{3})-(1-4dg_{0})a+2dg_{0}a^2
\right]^{2} \nonumber\\
&\times& 
\left[\{1-(g_{0}-g_{3})\}^{2}
-(g_{2}-g_{3})^{2} \right]^{d-2}.
\label{eqn:detmE}
\end{eqnarray}
It proves (\ref{eqn:P0}) of Theorem \ref{thm:P0andP00} (i).

It should be noted that, if we put 
$n=1$ and take $a \downarrow 0$ limit
in (\ref{eqn:detmE}), we have
the formula
$$
P_{0} =
\frac{4(d-1)^{2}}{d}
(1-d \bar{g}_{03}) \bar{g}_{03}^{2}
[(1-\bar{g}_{03})^2 -\bar{g}_{23}^2 ]^{d-2},
$$
where
$$
\bar{g}_{03}=\lim_{a \downarrow 0}(g_{0}-g_{3}),
\qquad 
\bar{g}_{23}=\lim_{a \downarrow 0}(g_{2}-g_{3}).
$$
In particular,
$\bar{g}_{03}=1/\pi$ 
and $\bar{g}_{23}=1-1/\pi$ 
for $d=2$ \cite{S64}, and thus
we have
$$
 P_{0} =\frac{2}{\pi^2} \left(1-\frac{2}{\pi} \right),
 \quad d=2.
$$
This coincides with the value of
$P_{0}$ obtained by Majumdar and Dhar \cite{MD91}
for the two-dimensional BTW model.

We can also find that the matrix $m^{*}(\lambda)$
defined by (\ref{eqn:mst}) is in the form 
(\ref{eqn:R1}) with
$$
\begin{array}{ll}
u=-2d, \qquad 
&b=(1-{\rm e}^{\lambda})/a^{1/2}, \cr
c=(1-{\rm e}^{-\lambda})/a^{1/2}, \quad
&q=(1-{\rm e}^{\lambda})/a^{1/2} 
-2da^{1/2} (g_{1}-{\rm e}^{\lambda}g_{0}), \cr
e=(1-{\rm e}^{-\lambda})/a^{1/2} 
-2da^{1/2} (g_{1}-{\rm e}^{-\lambda}g_{0}), \quad
&f=(g_{1}-g_{3})-{\rm e}^{\lambda}(g_{0}-g_{1}), \cr
s=(g_{1}-g_{3})-{\rm e}^{-\lambda}(g_{0}-g_{1}), \quad
&v=(g_{1}-g_{0})-{\rm e}^{\lambda}(g_{0}-g_{1}), \cr
k=(g_{1}-g_{0})-{\rm e}^{-\lambda}(g_{0}-g_{1}), \quad
&h=(g_{1}-g_{2})-{\rm e}^{\lambda}(g_{0}-g_{1}), \cr
t=(g_{1}-g_{2})-{\rm e}^{-\lambda}(g_{0}-g_{1}). &
\end{array}
$$
The formula (\ref{eqn:detR1}) gives
$$
\det m^{*}(\lambda)=
-2d \left[ \{1-(g_{0}-g_{3}) \}^{2}
-(g_{2}-g_{3})^2 \right]^{d-2}
\times \det S, 
$$
where
$$
\det S = b_{1}(d,a,\lambda)
+b_{2}(d,a,\lambda) \frac{1}{n}.
$$
with some functions $b_1$ and $b_2$
of $d, a, \lambda$.
Since (\ref{eqn:lambda}) gives
$$
  {\rm e}^{\lambda(a)} = 1+a+\sqrt{a(a+2)} 
  = 1+ \sqrt{2} a^{1/2}+\cO(a),
  \quad \mbox{as $a \downarrow 0$},
$$
we found that
\begin{eqnarray}
b_{1}(d,a,\lambda) &=& \cO(a^2),
\nonumber\\
b_{2}(d,a,\lambda) &=& \frac{4(d-1)}{d}
(g_{0}-g_{3})
\{1-d(g_{0}-g_{3})\}\{1+(d-1)(g_{0}-g_{3}) \}
+\cO(a^{1/2}),
\, \mbox{as $a \downarrow 0$}.
\nonumber
\end{eqnarray}
Thus we obtain
$$
\lim_{a \downarrow 0} \frac{\det m^{*}(\lambda)
\det m^{*}(-\lambda) }{(\det m)^2}
=\left[\frac{2d \{1+(d-1)\bar{g}_{03}\} }
{(d-1) \bar{g}_{03}} \right]^{2}.
$$
Since
$\lim_{a \downarrow 0} c_{1}(d,a)/a^{(d-3)/4}=
(d/(2 \pi^{2}))^{(d-3)/4}/(4 \pi)$,
(\ref{eqn:limA}) of Theorem \ref{thm:P0andP00} is proved.

\SSC{Discussions \label{sec:discussion}}
\subsection{Critical exponent $\nu_a$
\label{sec:exponent}}
The results (\ref{eqn:thm11}) of Theorem \ref{thm:main1}
and (\ref{eqn:thm21}) of Theorem \ref{thm:P0andP00} mean that
both of $G(\x(r))$ and $C_{00}(\x(r))$ decay exponentially
as increasing $r$ with a correlation length $\xi(d,a)$.
Since $\xi(d, a) < \infty$ for any $a > 0$, the stationary state 
of the DASM is non-critical \cite{TK00}.
Moreover the theorems imply that, if 
we make the parameter $n$ be large with a fixed $m$, 
then the value of $a=m/(2dn)$ can be small and 
\begin{eqnarray}
\label{eqn:hatG}
n G(\x(r)) &\simeq& c_{1}(d) a^{(d-3)/4}
\frac{{\rm e}^{-r/\xi(d,a)}}{r^{(d-1)/2}},
\\
\label{eqn:hatC}
C_{00}(\x(r)) &\simeq& c_{2}(d) a^{(d+1)/2}
\frac{{\rm e}^{-2r/\xi(d,a)}}{r^{d-1}}, \quad \mbox{as $r \uparrow \infty$},
\end{eqnarray}
where 
$c_{1}(d)=(d/ (2 \pi^{2}))^{(d-3)/4}/(4 \pi)$
and $c_{2}(d)$ is given by (\ref{eqn:limA}).

Consider a series of DASMs with increasing $n$ with a fixed $m$.
Then we will have an increasing series of correlation lengths $\{\xi(d,a)\}$
and we will see the asymptotic divergence, 
\begin{equation}
  \xi(d, a) \simeq \frac{1}{\sqrt{2d}} a^{-\nu_{a}} \quad 
  \hbox{as} \quad a \to 0
\label{eqn:result3}
\end{equation}
with
\begin{equation}
  \nu_{a}=\frac{1}{2} \quad \hbox{for all} \quad
d \geq 2.
\label{eqn:result4}
\end{equation}
We notice that, if we identify $a$ with a reduced temperature
\begin{equation}
t=\frac{|T-T_{{\rm c}}|}{T_{{\rm c}}}
\label{eqn:red_t}
\end{equation}
around a critical temperature $T_{{\rm c}}$ in the equilibrium
spin system, (\ref{eqn:hatG}) with
(\ref{eqn:result3}) and (\ref{eqn:result4}) is exactly in
the Ornstein-Zernike form of correlations
in the mean-field theory of equilibrium phase transitions
(see, for instance, Eq.(61) in Section 3.1 of \cite{ID89}). 
This implies that we can regard (\ref{eqn:result3})
as a critical phenomenon with a parameter $a$ approaching to its
critical value $a_{\rm c}=0$ and we can say that the associated 
{\it critical exponent} $\nu_{a}$ is exactly determined as 
(\ref{eqn:result4}).
Vanderzande and Daerden discussed the exponent
$\nu_{a}$ for the DASM on more general lattices \cite{VD01}.

This exponent may be identified with the critical exponent
$\nu=1/2$ obtained by Vespignani and Zapperi
by the generalized mean-field theory \cite{VZ98}.
They claimed that they made only use of conservation laws
to evaluate $\nu=1/2$ and thus
at least on this result their mean-field theory is
exact for any $d \geq 2$.
The present work justifies their conjecture.
We can conclude that with respect to the avalanche
propagators and height-$(0,0)$ correlation functions
the upper critical dimension of the ASM is two.
This result does not contradict to the result
by Priezzhev \cite{P00},
since he studied the intersection phenomena of
avalanches and for them the upper critical dimension is four.

The results (\ref{eqn:hatG}) and (\ref{eqn:hatC})
suggest that there exists a scaling limit such that
\begin{eqnarray}
  && \lim_{\substack{r \uparrow \infty, a \downarrow 0: \cr
  a^{1/2} r=\kappa/\sqrt{2d} }}
  r^{d-2} n G(\x(r)) = \cF_G(\kappa), 
\nonumber\\
  && \lim_{\substack{r \uparrow \infty, a \downarrow 0: \cr
  a^{1/2} r=\kappa/\sqrt{2d} }}
  r^{2d} C_{00}(\x(r)) = \cF_C(\kappa), \quad 0 < \kappa < \infty
\nonumber
\end{eqnarray}
with
\begin{eqnarray}
 \cF_G(\kappa) &=& 2^{-(d+1)/2} \pi^{-(d-1)/2} \kappa^{(d-3)/2} e^{-\kappa},
\nonumber\\
 \cF_C(\kappa) &=& 2^{-(d+1)} \pi^{-(d-1)}
\left[ \frac{1+(d-1)\bar{\gamma}}{(d-1)\bar{\gamma}} \right]^2 
\kappa^{d+1} {\rm e}^{-\kappa}, 
\nonumber
\end{eqnarray}
This observation is consistent with the statement
\begin{equation}
G(\x(r)) \sim r^{-(d-2)}, \quad
\mbox{as $r \uparrow \infty$}
\label{eqn:MD4}
\end{equation}
and (\ref{eqn:MD3})
claimed by Majumdar and Dhar \cite{MD91} for the 
self-organized criticality realized in the 
$d$-dimensional BTW model with $d \geq 2$.
(Note that for the two-dimensional BTW model,
$G(\x(r))-G(0) \simeq - (1/2\pi) \log r$,
as $r \uparrow \infty$.)

\subsection{The $q \to 0$ limit of the Potts model
\label{sec:Potts}}
Majumdar and Dhar \cite{MD92} discussed the relationship
between the ASM and the $q \downarrow 0$ limit of the
{\it $q$-state Potts model}.
For $q \in \{2,3, \dots\}$, the $q$-state Potts model
on the lattice $G_L=(G_L^{(v)}, G_L^{(e)})$
given by Definition \ref{thm:graph} is defined as follows.
At each vertex $\v \in G_L^{(v)} = \Lambda_L \cup \{\r\}$,
put a spin variable 
$s(\x) \in \{1,2, \dots, q\}$. 
The Hamiltonian for the configuration 
$s=\{s(\v)\}_{\v \in G_L^{(v)}}$ is given by
$$
\cH(s)=- \sum_{e=\{\v, \w\} \in G_L^{(e)}}
\1(s(\v)=s(\w)).
$$
The partition function of the Potts model in the Gibbs ensemble
with a temperature $T>0$ is defined by
\begin{eqnarray}
Z(q, T) &=& \sum_{s \in \{1,2, \dots, q\}^{G_L^{(v)}}} e^{-\cH(s)/T}
\nonumber\\
&=& \sum_{s \in \{1,2 \dots, q \}^{G_L^{(v)} } } 
\prod_{e=\{\v, \w\} \in G_L^{(e)} }
\Big[1+ \chi \1(s(\v)=s(\w)) \Big]
\label{eqn:Z1}
\end{eqnarray}
with $\chi=e^{1/T}-1$.
We consider a subset of $G_L^{(e)}$ denoted by $E \subset G_L^{(e)}$.
Each connected component in $E$ is called a cluster.
Let $c(E)$ be the number of disconnected clusters of $E$;
$E=\bigcup_{i=1}^{c(E)} E_i$, where
$E_i \cap E_j = \emptyset, i \not= j$.
If a vertex $\v \in G_L^{(v)}$ is not connected by any edge in $E$,
we write $\v \notin E$.
By performing binomial expansions and taking the summation
over spin configurations in (\ref{eqn:Z1}), we obtain the
Fortuin-Kasteleyn representation of partition function,
\begin{equation}
Z(q,T)=\sum_{E \subset G_L^{(e)}}
q^{|\{\v \in G_L^{(v)} ; \v \notin E\}|}
q^{c(E)} \chi^{|E|},
\label{eqn:Z2}
\end{equation}
where $|E|$ denotes the number of edges in $E$.
Note that we can regard (\ref{eqn:Z2}) as a function of
$q \in \R$ and $T >0$.
We consider the asymptotics of (\ref{eqn:Z2}) in the limit $q \downarrow 0$.
The dominant terms in this limit should be
with $E$ such that $c(E)=1$ and
$\{\v \in G_L^{(v)}: \v \notin E\}=\emptyset$
$\Longleftrightarrow E$ contains all vertices in $G_L^{(v)}$
$\Longleftrightarrow E$ is a spanning subgraph of $G_L$.
If we further take the high-temperature limit $T \uparrow \infty$
$\Longleftrightarrow \chi \downarrow 0$,
we have only spanning subgraphs with a minimal number of edges,
which are just the spanning trees. 
Then we have
$$
\lim_{T \uparrow \infty} \lim_{q \downarrow 0}
T^{(2L+1)^d} q^{-1} Z(q,T)=|\cT_L|,
$$
where $\cT_L$ is the collection of all spanning trees
on $G_L$.
As shown in Section \ref{sec:allowed},
there establishes a bijection between $\cT_L$ and $\cA_L$
(Lemma \ref{thm:AandT}) and
$\cA_L=\cR_L$ (Proposition \ref{thm:RandA_2}).
(The relation between the $q \downarrow 0$ limit of the
$q$-state Potts model with finite temperatures
and the ASM is discussed in Section 7.2 in \cite{Dhar06}.)
The two-dimensional $q$-state Potts model shows 
a continuous phase transition associated with {\it critical phenomena}
at a finite temperature $0< T_{\rm c} < \infty$
without external magnetic field $B=0$,
when $q=2,3$ and 4 \cite{Wu82}.

Usual critical phenomena of spin models are specified by
the behavior of two-point correlation functions
for the energy density $G_{\epsilon}(r, t, b, L)$
and for the order-parameter density $G_{\sigma}(r, t, b, L)$.
Here $r$ denotes the distance of two points,
$t$ the reduced temperature (\ref{eqn:red_t}), 
$b$ the reduced external field
$$
b=\frac{|B|}{T_{\rm c}},
$$
and $L$ the size of the lattice on which the model is defined.
It is conjectured in the scaling theory that,
if $L$ is sufficiently large and we observe the system
in the very vicinity of the critical point;
$t \ll 1, b \ll 1$, 
the correlation functions behave as
\begin{eqnarray}
G_{\epsilon}(r, t, b, L)
=L^{2 x_{\epsilon}} \cF_{\epsilon}
\left( \frac{r}{L}, t L^{y_t}, b L^{y_b} \right),
\nonumber\\
G_{\sigma}(r, t, b, L)
=L^{2 x_{\sigma}} \cF_{\sigma}
\left( \frac{r}{L}, t L^{y_t}, b L^{y_b} \right),
\label{eqn:scaling}
\end{eqnarray}
with the scaling exponents $x_{\epsilon}, x_{\sigma}$,
$y_{\epsilon}, y_{\sigma}$,
and the scaling functions $\cF_{\epsilon}, \cF_{\sigma}$.
If the system is of $d$-dimensional, the hyperscaling relations
$x_{\epsilon} + y_{t}=d, x_{\sigma}+y_{b}=d$ hold 
(see, for instance, \cite{Hen99,ID89}).
From the scaling forms (\ref{eqn:scaling}),
we expect the power-law behavior of correlation functions
at the critical point ($t=b=0, L \uparrow \infty$)
such that
$$
G_{\epsilon}(r) \sim r^{-2 x_{\epsilon}},
\quad
G_{\sigma}(t) \sim r^{-2 x_{\sigma}},
\quad \mbox{as $r \uparrow \infty$},
$$
and in the off-critical regions with $L \uparrow \infty$,
the correlation length $\xi=\xi(t, b)$ behaves as
\begin{eqnarray}
&&\xi(t, 0) \sim t^{-\nu_t}
\quad \mbox{with} \quad \nu_{t}=\frac{1}{y_t},
\nonumber\\
&&\xi(0, b) \sim b^{-\nu_b}
\quad \mbox{with} \quad \nu_{b}=\frac{1}{y_b},
\quad \mbox{as $t \downarrow 0, b \downarrow 0$}.
\nonumber
\end{eqnarray}

For the two-dimensional $q$-state Potts model,
the critical exponents are determined as functions of $q$
through the parameter
$$
u=u(q)=\frac{2}{\pi} \cos^{-1} \left( \frac{\sqrt{q}}{2} \right)
$$
as \cite{Wu82}
$$
\begin{array}{ll}
\displaystyle{
x_{\epsilon}=\frac{1+u}{2-u}
}, &
\qquad
\displaystyle{
y_t=2-x_{\epsilon}=\frac{3(1-u)}{2-u},
} \cr
\displaystyle{
x_{\sigma}=\frac{1-u^2}{4(2-u)}
}, &
\qquad
\displaystyle{
y_b=2-x_{\sigma}=\frac{3(1-u)(5-u)}{4(2-u)}.
}
\end{array}
$$
They give the limits
$$
x_{\epsilon} \to 2, \quad
y_t \to 0, \quad
x_{\sigma} \to 0, \quad
y_b \to 2,
\quad \mbox{as $q \downarrow 0
\Longleftrightarrow u \uparrow 1$}.
$$
Majumdar and Dhar \cite{MD92} noted by their results
(\ref{eqn:MD3}) and (\ref{eqn:MD4}) for the BTW models that
the avalanche propagator $G(\x(r))$ and the height-$(0,0)$ correlation
function $C_{00}(\x(r))$ in ASM play the roles
of the order-parameter density correlation function $G_{\sigma}(r)$
and the energy density correlation function $G_{\epsilon}(r)$
in the critical phenomena, respectively.
In particular, in the two-dimensional case,
the power-law exponents are respectively given as
$$
2 x_{\sigma} \Big|_{q \downarrow 0} = 0 = d-2 \Big|_{d=2},
\qquad
2 x_{\epsilon} \Big|_{q \downarrow 0} = 4 = 2 d \Big|_{d=2}.
$$
Our interpretation of the present result (\ref{eqn:result4}) 
is that introduction of dissipation to the ASM
may correspond to imposing an external magnetic field $B$
to the Potts models and hence
$\nu_a=1/2$ is identified with
$$
\nu_b \Big|_{q \downarrow 0}
=\left. \frac{1}{y_b} \right|_{q \downarrow 0} = \frac{1}{2}.
$$

We remark that the critical exponents
for the specific heat $\alpha$,
for the order parameter $\beta$,
and for the magnetic-field susceptibility $\gamma$
of the$$
\alpha=\frac{2(1-2u)}{3(1-u)} \to -\infty, \quad
\beta=\frac{1+u}{12} \to \frac{1}{6}, \quad
\gamma=\frac{7-4u+u^2}{6(1-u)} \to \infty, 
\quad \mbox{as $q \downarrow 0
\Longleftrightarrow u \uparrow 1$}.
$$
We suspect some interpretation of the
value $\beta|_{q \downarrow 0}=1/6$ in the DASM.

\subsection{Recent topics on height correlations
\label{sec:toppics}}

In Section \ref{sec:correlation}  the one-point and
the two-point correlations of height-0 sites
were calculated for the DASM with general $d \geq 2$.
In the two-dimensional case, the three-point
and the four-point correlations were also
calculated for height-0 sites and general property
of `the height-0 field of ASMs' have been
extensively studied from the view point of 
a $c=-2$ conformal field theory
\cite{MR01,Dur09}. 

For the two-dimensional BTW model, in which the values
of stable height of sandpile are $h=0,1,2,$ and 3,
the height correlations have been calculated also for
$h \geq 1$.
Priezzhev determined $P_{\alpha}$ for
$\alpha \in \{0,1,2,3\}$, where the results
with $\alpha \geq 1$ are expressed using
multivariate integrals of determinantal integrands \cite{P94}.
Poghosyan et al. \cite{PGPR10} claimed that
the height-0 state is the only one showing pure
power-law-correlations and that
general form of height correlations for $h \geq 1$
contains logarithmic functions.
They showed that for $\alpha \geq 1$
$$
C_{0\alpha}(\x(r))
=\frac{P_{0\alpha}(\x(r))-P_0 P_{\alpha}}{P_0 P_{\alpha}}
\simeq \frac{1}{r^4}(c_1 \log r + c_2),
\quad \mbox{as $r \uparrow \infty$}
$$
with some constants $c_1, c_2$.
Moreover, they predicted that
$C_{\alpha \beta}(\x(r)) \sim \log^2r/r^4$
if $\alpha \geq 1$ and $\beta \geq 1$.
These results are discussed with the 
logarithmic conformal field theory.
See also \cite{GNPP13}.
We will see a lot of interesting open problems
concerning height correlations for the
BTW models and the DASMs in higher dimensions.

\vskip 0.3cm
\noindent{\bf Acknowledgements} \quad
This manuscript was prepared for the
workshop ``Probabilistic models with determinantal structure'',
(April 30th and May 1st, 2015) held at the
Faculty of Mathematics -- Institute of Mathematics for Industry,
Ito Campus, Kyushu University.
The present author would like to thank T. Shirai
for the invitation and for his hospitality.
He thanks T. Shirai, E. Verbitskiy, and T. Hara for useful discussions
in the workshop.
This work is supported in part by
the Grant-in-Aid for Scientific Research (C)
(No.26400405) of Japan Society for
the Promotion of Science.


\end{document}